\documentclass[aps,pra,showpacs,twocolumn,10pt]{revtex4-1}
\usepackage{graphicx}
\usepackage{appendix}
\usepackage{amssymb}
\usepackage{amsmath}
\usepackage{bbm}
\usepackage{hyperref}
\usepackage{color,soul}
\usepackage[bottom]{footmisc}
\usepackage{floatrow}
\usepackage[labelformat=parens,caption=false]{subfig}

\begin{document}

\title{Charge-spin response and collective excitations in Weyl semimetals}

\author{Sayandip Ghosh}
\author{Carsten Timm} 
\email{carsten.timm@tu-dresden.de}
\affiliation{Institute of Theoretical Physics, Technische Universit{\"a}t Dresden, 01062 Dresden, Germany}

\date{October 30, 2018}

\begin{abstract}
Weyl semimetals are characterized by unconventional electromagnetic response. We present analytical expressions for all components of the frequency- and wave-vector-dependent charge-spin linear-response tensor of Weyl fermions. The spin-momentum locking of the Weyl Hamiltonian leads to a coupling between charge and longitudinal spin fluctuations, while transverse spin fluctuations remain decoupled from the charge. A real Weyl semimetal with multiple Weyl nodes can show this charge-spin coupling in equilibrium if its crystal symmetry is sufficiently low. All Weyl semimetals are expected to show this coupling if they are driven into a non-equilibrium stationary state with different occupations of Weyl nodes, for example by exploiting the chiral anomaly. Based on the response tensor, we investigate the low-energy collective excitations of interacting Weyl fermions. For a local Hubbard interaction, the charge-spin coupling leads to a dramatic change of the zero-sound dispersion: its velocity becomes independent of the interaction strength and the chemical potential and is given solely by the Fermi velocity. In the presence of long-range Coulomb interactions, the coupling transforms the plasmon modes into spin plasmons.  For real Weyl semimetals with multiple Weyl nodes, the collective modes are strongly affected by the presence of parallel static electric and magnetic fields, due to the chiral anomaly. In particular, the zero-sound frequency at fixed momentum and the spin content of the spin plasmons go through cusp singularities as the chemical potential of one of the Weyl cones is tuned through the Weyl node. We discuss possible experiments that could provide smoking-gun evidence for Weyl physics.
\end{abstract}

\maketitle

\section{Introduction}
\label{sec1}

Within the large field of topological condensed matter physics, the study of Weyl semimetals (WSMs) \cite{Burkov2011,Vafek2014,Chiu2016,Yan2017,Burkov2018,Armitage2018} has received a strong boost by the discovery of several candidate materials \cite{Xu2015,Lv2015a,Huang2015,Lv2015c,Yang2015,Xu2015a,Nakatsuji2015,Xiong2015,Liu2016,Xu2016,Hirschberger2016, Nayak2016,Shekhar2016,Ikhlas2017,Li2017,Liu2018,Wang2018,Hutt2018}. In WSMs, non-degenerate bands touch at points in momentum space, called ``Weyl nodes.'' This requires time-reversal or spatial inversion symmetry to be broken since otherwise all bands would be spin degenerate. In magnetic WSMs such as the candidates $\mathrm{Mn}_3(\mathrm{Ge},\mathrm{Sn})$ \cite{Nakatsuji2015,Nayak2016,Ikhlas2017,Li2017,Yang2017,Ito2017} and $\mathrm{Co}_3\mathrm{Sn}_2\mathrm{S}_2$ \cite{Liu2018,Wang2018,Xu2018} as well as $\mathrm{Na}_3\mathrm{Bi}$ in a magnetic field \cite{Xiong2015} and $(\mathrm{Gd},\mathrm{Nd})\mathrm{PtBi}$ in a magnetic field \cite{Hirschberger2016,Shekhar2016,Hutt2018}, time-reversal symmetry (TRS) is broken. On the other hand, time-reversal-symmetric WSMs with broken inversion symmetry have been realized in the family $(\mathrm{Ta},\mathrm{Nb})(\mathrm{As},\mathrm{P})$~\cite{Xu2015,Lv2015a,Huang2015,Lv2015c,Yang2015,Xu2015a,Liu2016,Xu2016}.

In WSMs, Weyl nodes act as sources and sinks of Berry curvature (of, say, the lower band), which is analogous to a magnetic field in momentum space. The corresponding magnetic monopole charge, or ``chirality,'' is given by the flux of the Berry curvature over a Fermi surface enclosing the node and is quantized to an integer value. Since the net monopole charge summed over all Weyl nodes in the Brillouin zone must be zero, Weyl nodes always appear in pairs of opposite chirality \cite{Nielsen1981a,Nielsen1981b,Nielsen1981}. Near each node, the band dispersion is linear and the effective Hamiltonian has the form of the well-known Weyl Hamiltonian \cite{Weyl1929,Peskin1995}. The corresponding massless low-energy quasiparticles, the Weyl fermions, exhibit a chiral or Adler-Bell-Jackiw anomaly \cite{Adler1969,Bell1969,Nielsen1983}, which means that, in the presence of parallel electric and magnetic fields, the number of fermions close to Weyl nodes of opposite chirality is not conserved separately. WSMs show a plethora of exotic optical and transport properties, some, but not all, of which are caused by the chiral anomaly. Examples are an anomalous Hall effect in WSMs with broken time-reversal symmetry \cite{Goswami2013}, a chiral magnetic effect, i.e., a dynamical current parallel to the magnetic field in WSMs with broken inversion symmetry \cite{Ma2015,Zhong2016,Burkov2018}, and a negative magnetoresistance for the magnetic field parallel to the electric field~\cite{Kim2013}. Moreover, the chiral anomaly leads to the appearance of a term proportional to $\mathbf{E}\cdot\mathbf{B}$ in the electromagnetic action, with a non-uniform prefactor \cite{Zyuzin2012}. This term implies a magnetoelectric effect: an applied electric field generates a magnetization, while an applied magnetic field generates an electric polarization \cite{Vafek2014,Burkov2018,Armitage2018}. The magnetoelectric effect is encoded in the coupled charge-spin linear-response tensor, the frequency and wave-vector dependence of which has not been calculated completely so far.

In this paper, we investigate the response of Weyl fermions to time- and space-dependent electric and magnetic perturbations by calculating all components of the composite charge-spin linear-response tensor for a single Weyl node in Sec.\ \ref{sec2}. Based on this, we investigate observable manifestations for real Weyl semimetals with pairs of Weyl nodes with opposite chirality in Sec.\ \ref{sec3}. After briefly discussing the electromagnetic response, we present a detailed study of the collective excitations within the random phase approximation, exploring in particular the impact of the coupling between density and spin excitations. The chiral anomaly significantly affects the excitation modes, which can be probed by optical pump-probe experiments. Finally, we summarize our results in Sec.~\ref{sec4}.

\section{Linear response}
\label{sec2}

We start from an effective Hamiltonian describing the low-energy physics of Weyl fermions in the vicinity of a Weyl node residing at ${\bf Q}$ in the Brillouin zone,
\begin{equation}
{\cal H} = \chi  v_F {\bf k} \cdot{\boldsymbol \sigma} - \mu_\chi ,
\label{Hamiltonian}
\end{equation}
where ${\bf k}$ is the momentum relative to the Weyl node, ${\boldsymbol \sigma}$ is the vector of Pauli matrices representing the electron spin, $\chi = \pm 1$ denotes the chirality describing the relative orientation of spin and momentum, and $\mu_\chi$ is the chirality-dependent chemical potential. We set $\hslash = 1$ and assume isotropic Fermi velocity $v_F$. However, our results are qualitatively insensitive to the anisotropy.

The first-order response to electromagnetic perturbations is described by a $4 \times 4$ linear-response tensor, the components of which are determined by correlation functions,
\begin{equation}
\Pi_{\alpha \alpha'}({\bf q},i\omega_n)
  = \frac{1}{N} \int_{0}^{\beta} d\tau\, e^{i\omega_n\tau} \langle T_{\tau}
  \alpha({\bf q},{\tau}) \alpha'(-{\bf q},0) \rangle ,
\label{Pi0}
\end{equation}
where $\alpha,\alpha' \in \{\rho, \sigma_l\}$ refer to the Fourier-transformed density and spin operators defined as $\rho({\bf q}) = \sum_{{\bf k}\sigma} c_{{\bf k} + {\bf q},\sigma}^\dagger c_{{\bf k},\sigma}$ and $\sigma_l({\bf q}) = \sum_{{\bf k}\zeta\zeta'} c_{{\bf k} + {\bf q},\zeta}^\dagger \sigma^l_{\zeta,\zeta'} c_{{\bf k},\zeta'}$, respectively, in terms of the fermionic annihilation (creation) operators $c_{{\bf k},\sigma}$ ($c_{{\bf k},\sigma}^\dagger$). $\beta=1/k_BT$ is the inverse temperature, $i\omega_n$ are Matsubara frequencies, and $T_\tau$ is the time-ordering directive in imaginary time. All response functions can be written as sums of interband (denoted by superscript $-$) and intraband (superscript $+$) contributions: $\Pi_{\alpha\alpha'} = \Pi^-_{\alpha\alpha'} + \Pi^+_{\alpha\alpha'}$. The retarded response functions are obtained by analytic continuation $i\omega_n \to \omega + i\delta$.

\subsection{Separate charge and spin responses}
\label{section_cc_ss}

The retarded charge response functions $\Pi^\mp_{\rho\rho}(\mathbf{q},\omega)$ were obtained by Lv and Zhang \cite{Lv2013} (see also Ref.\ \cite{Zhou2015}). At zero temperature they are given by
\begin{align}
\Pi^\mp_{\rho\rho}({\bf q},\omega)
  &= \frac{1}{2N} \sum_{\bf k} \bigg(1 \mp \frac{\mathbf{k}'\cdot\mathbf{k}}{k' k} \bigg) \nonumber \\
&{}\times \bigg[ \frac{1}{v_F (k \pm k') - \omega - i \delta} + (\omega \rightarrow - \omega) \bigg] ,
\label{Pi1}
\end{align}
where ${\bf k'} = {\bf k} + {\bf q}$, $k = |{\bf k}|$ etc. For completeness, the explicit form of the charge response tensor $\Pi_{\rho\rho}({\bf q},\omega)$ is presented in Appendix \ref{app_charge}. It only depends on the magnitude of the wave vector ${\bf q}$, as expected for the isotropic Hamiltonian~(\ref{Hamiltonian}). The charge response is even in chirality.

The spin response tensors contain diagonal and off-diagonal components in spin space. The diagonal terms
\begin{align}
\Pi^\mp_{\sigma_l\sigma_l} ({\bf q},\omega)
  &= \frac{1}{2N} \sum_{\bf k} \bigg(1 \pm \frac{ k'_m k_m + k'_n k_n - k'_l k_l}{k' k} \bigg) \nonumber \\
&{}\times \bigg[ \frac{1}{v_F (k \pm k') - \omega - i \delta} + (\omega \rightarrow - \omega) \bigg] ,
\label{Pi2}
\end{align}
where $l$, $m$, $n$ refer to three orthogonal coordinate axes with $\varepsilon_{lmn} = +1$, have recently been evaluated by Thakur \textit{et al.}\ \cite{Thakur2018} and Zhou and Chang \cite{Zhou2018}. These authors are mainly interested in the \emph{current} response, which is, however, closely related to the spin response due to spin-momentum locking. On the other hand, the off-diagonal components read as
\begin{align}
\Pi^\mp_{\sigma_l\sigma_m}({\bf q},\omega)
  &= \frac{1}{2N} \sum_{\bf k} \left[ \frac{ \mp \frac{k_l k'_m + k'_l k_m}{k k'} + i\,\big(\frac{k'_n}{k'}
  \pm \frac{k_n}{k}\big)}{v_F ( k \pm k') - \omega - i \delta} \right. \nonumber \\
&{}+ \left. \frac{\mp \frac{k_l k'_m + k'_l k_m}{k k'} - i\,\big(\frac{k'_n}{k'} \pm \frac{k_n}{k}\big)}
  {v_F (k \pm k') + \omega + i \delta} \right]. 
\label{Pi3}
\end{align}
Evidently, the spin response depends on both the magnitude and the direction of ${\bf q}$. We consider ${\bf q}$ along the positive \textit{z} axis, without loss of generality. The response tensor for $\mathbf{q}$ in other directions can then be obtained by simultaneously rotating $\mathbf{q}$ as a vector and $\Pi^\mp_{\sigma_l\sigma_m}$ as a second-rank tensor with indices $l,m$. The diagonal terms have longitudinal ($\Pi^\mp_{\sigma_z \sigma_z}$) and transverse ($\Pi^\mp_{\sigma_x \sigma_x} = \Pi^\mp_{\sigma_y \sigma_y}$) components, which differ in the orientation between the wave vector and the spin. The imaginary parts of the longitudinal and transverse diagonal spin responses are related to the charge response by multiplicative factors \cite{Thakur2018}. For the real parts, the situation is more complicated since they contain the unphysical cutoff-dependent term ${\Lambda^2}/{6 \pi^2 v_F}$, where $\Lambda$ is the ultraviolet cutoff for the momentum sum \cite{Polini2009,Kargarian2015,Thakur2018,Zhou2018,Ominato2018}. This term arises from the presence of the infinite sea of negative-energy states for the effective Weyl Hamiltonian. The spin response function should be regularized by taking the $\mu = 0$ ground state as the reference system \cite{Sabio2008,Zhou2018,Polini2009,Thakur2018}, which corresponds to subtracting the $\Lambda^2$ term from the spin response tensor. Details of the calculation are given in Appendix \ref{app_spin}, where we also reproduce the explicit results for the diagonal components~\cite{Thakur2018}.

Among the off-diagonal terms, only the transverse components $\Pi^\mp_{\sigma_x \sigma_y} = -\Pi^\mp_{\sigma_y\sigma_x}$ differ from zero for ${\bf q} = q\hat{\bf z}$. Hence, the transverse and longitudinal components of the spin response decouple, which can be attributed to the rotational invariance of the model about the wave vector $\mathbf{q}$. By rewriting the momentum sum in Eq.\ (\ref{Pi3}) as an integral and performing the angular part, we obtain
\begin{align}
&\Pi^\mp_{\sigma_x \sigma_y}(q \hat{\mathbf{z}},\omega)
  = \frac{i}{16 \pi^2 q^2}\, \int dk \int dk'\, (k \mp k') \nonumber \\
&{}\times [q^2 - (k \pm k')^2]
  \left[ \frac{1}{v_F (k \pm k') - \omega - i \delta}
  - (\omega \rightarrow - \omega) \right] .
\end{align}
The integrals can be evaluated explicitly. The results are given in Appendix \ref{app_spin}. In the undoped case, the response function becomes simple and purely imaginary:
\begin{equation}
\Pi_{\sigma_x \sigma_y}^{\rm in}(q \hat{\mathbf{z}},\omega) = i\, \frac{q \omega}{24 \pi^2 v_F^2} .
\end{equation}
The usually considered off-diagonal response function $\Pi_{\sigma_+\sigma_-}$ with $\sigma_\pm = \sigma_x \pm i \sigma_y$ is related to the calculated components by
\begin{equation}
\Pi_{\sigma_+\sigma_-}(q \hat{\mathbf{z}},\omega)
  = 2\, \Pi_{\sigma_x\sigma_x}(q \hat{\mathbf{z}},\omega) - 2i\, \Pi_{\sigma_x\sigma_y}(q \hat{\mathbf{z}},\omega) ,
\end{equation}
where symmetries have been used. Hence, the real part of $\Pi_{\sigma_+\sigma_-}$ is related to the imaginary part of $\Pi_{\sigma_x\sigma_y}$.

Note that the Hall conductivity is proportional to the off-diagonal current-current correlation function, which can be expressed in terms of the spin response by~\cite{Raghu2010,Thakur2018}
\begin{equation}
\sigma_{xy}(\mathbf{q},\omega)
  = \frac{i e^2 v_F^2}{\omega}\, \Pi_{\sigma_x \sigma_y}(\mathbf{q},\omega) .
\label{Hall1}
\end{equation}
In the static and uniform limit, the Hall conductivity is known to be proportional to the separation of Weyl nodes \cite{Yang2011,Burkov2011b,Zyuzin2012,Kargarian2015}. This relies on the observation that the Hall conductivity contributed by two-dimensional slices of momentum space jumps by a quantized amount when the slice passes a Weyl node. There is a corresponding jump in the off-diagonal transverse spin response, which is included in our results. However, the absolute value of the Hall conductivity and hence of the off-diagonal transverse spin response function cannot be determined from a continuum field theory \cite{Burkov2011b} (see also Appendix \ref{app_spin}). Since the off-diagonal transverse spin response is not central for this paper, we do not pursue this issue here.

\begin{figure}[tb]
\includegraphics[width=0.8\columnwidth]{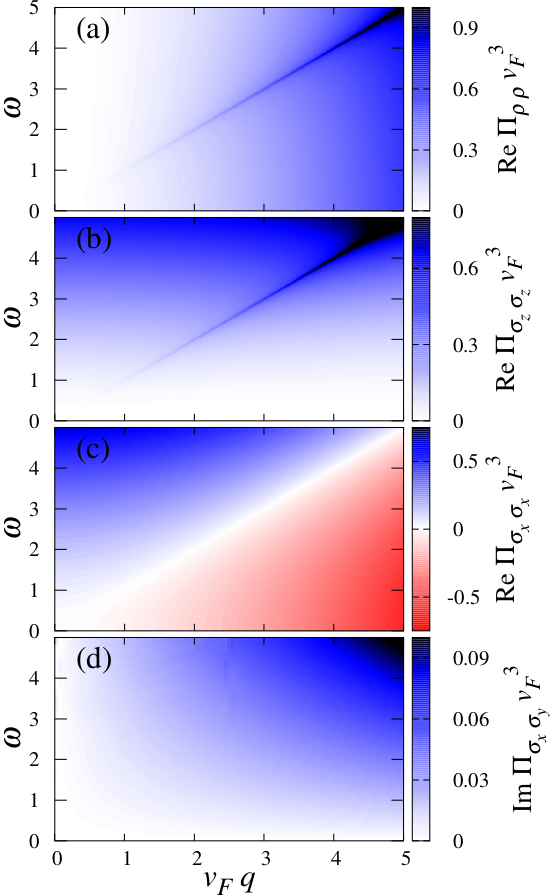}
\caption{Charge and spin responses of an undoped ($\mu=0$) Weyl cone: wave-vector and frequency dependence of (a) $\mathrm{Re}\,\Pi_{\rho \rho}$, (b) $\mathrm{Re}\,\Pi_{\sigma_z \sigma_z}$, (c) $\mathrm{Re}\,\Pi_{\sigma_x \sigma_x}$, and (d) $\mathrm{Im}\,\Pi_{\sigma_x \sigma_y}$ for ${\bf q} = q \hat{\mathbf{z}}$. As discussed in the text, $\mathrm{Im}\,\Pi_{\sigma_x \sigma_y}$ is related to the off-diagonal response function $\mathrm{Re}\,\Pi_{\sigma_+\sigma_-}$, whereas $\mathrm{Re}\,\Pi_{\sigma_x \sigma_y}$ vanishes in the undoped case. Here, the ultraviolet cutoff is taken to be $v_F \Lambda = 50$.}
\label{fig_res} 
\end{figure}

For illustration, the wave-vector and frequency dependence of the charge and spin response functions is presented in Fig.\ \ref{fig_res} for a single, undoped Weyl cone. The ultraviolet cutoff is taken as $v_F \Lambda = 50$ throughout the paper. For a typical bandwidth of $1\,\mathrm{eV}$, the corresponding frequency scale is $20\,\mathrm{meV}$ and the wave-vector scale is $10^6\,\mathrm{cm}^{-1}$ for $v_F \approx 10^6\,\mathrm{ms}^{-1}$.

Figures \ref{fig_res}(a)--\ref{fig_res}(c) depict the real parts of the charge and diagonal spin response functions. The charge and longitudinal spin response exhibits a logarithmic singularity at $\omega=v_Fq$ from the joint density of states. The singularity would be broadened by deviations from linear dispersion. In the diagonal transverse spin response [Fig.\ \ref{fig_res}(c)] the logarithmic singularity is pushed to the first derivative by a vanishing prefactor $\omega^2-v_F^2 q^2$. The origin of this suppression is that the ``on-shell'' transverse response is prohibited by spin-momentum locking. The imaginary parts are not plotted here since they are simple rational functions of $q$ and $\omega$ multiplied by the step function $\theta(\omega-v_Fq)$ [see Eqs.\ (\ref{IPirhorhoin}), (\ref{Pizzin}), and (\ref{Pixxin})].

Figure \ref{fig_res}(d) shows the imaginary part of the off-diagonal transverse spin response function $\Pi_{\sigma_x\sigma_y}$, which, as noted above, is related to the real part of the usual off-diagonal spin response function $\Pi_{\sigma_+\sigma_-}$. Unlike the components plotted in Figs.\ \ref{fig_res}(a)--\ref{fig_res}(c), it is a real analytic function of $q$ and $\omega$ and does not show any singularity at the Weyl-fermion dispersion $\omega=v_F q$. However, $\Pi_{\sigma_x\sigma_y}$ being nonzero relies on the chirality of the Weyl Hamiltonian.

\subsection{Coupled charge-spin response}
\label{section_cs}

For free electrons in a periodic potential, the linear charge and spin responses are decoupled. In real materials, their coupling, i.e., a magnetoelectric effect \cite{Fiebig2005}, can be caused by spin-orbit interaction or coupling to strain but is usually weak. In WSMs, spin is locked to momentum, which leads to strong correlations of spin and current and hence to a large magnetoelectric effect \cite{Vafek2014,Burkov2018,Armitage2018}. Consequently, the coupled charge-spin response function $\Pi_{\rho \sigma_l}$ is large, as we show in the following. Details of the calculation are relegated to Appendix \ref{app_coupled}. We find that only the longitudinal contribution, for which the spin direction is parallel to the wave vector, is nonzero. Furthermore, it satisfies $\Pi_{\rho\sigma_l}=\Pi_{\sigma_l\rho}$. The longitudinal contribution can be decomposed into two parts, an \emph{intrinsic} one which is nonzero already in the undoped case and an \emph{extrinsic} one that only emerges upon doping.

For the intrinsic case, the Fermi level lies at the Weyl node, and only interband transitions from the completely filled valence band to the empty conduction band contribute. The imaginary and real parts of the intrinsic charge-spin response are given by
\begin{align}
{\rm Im}\,\Pi_{\rho \sigma_z}^{\rm in}(q \hat{\mathbf{z}},\omega) &= \frac{q \omega}{24 \pi v_F^2}\,
  \theta(\omega - v_F q) , \\
{\rm Re}\,\Pi_{\rho \sigma_z}^{\rm in}(q \hat{\mathbf{z}},\omega) &= \frac{q\omega}{24 \pi^2 v_F^2}\,
  \ln \left| \frac{4 v_F^2 \Lambda^2}{v_F^2 q^2 - \omega^2} \right| ,
\label{Pi4}
\end{align}
where $\theta(x)$ is the Heaviside step function. In the presence of doping, the charge-spin response depends only on the magnitude of the chemical potential $\mu$, due to the particle-hole symmetry of the Weyl Hamiltonian. For electron doping, $\mu = v_F k_F > 0$, the imaginary and real parts of the extrinsic contribution read as
\begin{widetext}
\begin{align}
{\rm Im}\,\Pi^{\rm ex}_{\rho \sigma_z}(q \hat{\mathbf{z}},\omega)
  &= \frac{\omega}{8 \pi q v_F^2} \bigg[ \theta(v_F q - \omega) \big( [\alpha(q,\omega)
  - \alpha(q,-\omega)] \theta(2\mu - v_F q - \omega)
  + \alpha(q,\omega)  \theta(2\mu - v_F q + \omega)\theta(v_F q + \omega - 2\mu)\big) \nonumber \\
& {}+  \theta(\omega - v_F q) \bigg( {-}\alpha(-q,-\omega)
  \theta(2\mu + v_F q - \omega)\theta(v_F q + \omega - 2\mu)
  - \frac{q^2}{3}\, \theta(2\mu - v_F q - \omega) \bigg) \bigg] ,
\label{Pi5} \\
{\rm Re}\,\Pi^{\rm ex}_{\rho \sigma_z}(q \hat{\mathbf{z}},\omega)
  &= \frac{\omega}{8\pi^2 v_F^2 q} \bigg[ \frac{8\mu^2}{3v_F^2} - \alpha(q,\omega)\beta(q,\omega)
  - \alpha(-q,\omega)\beta(-q,\omega) - \alpha(q,-\omega)\beta(q,-\omega)
  - \alpha(-q,-\omega)\beta(-q,-\omega) \bigg] ,
\label{Pi6}
\end{align}
\end{widetext} 
where we have defined
\begin{equation}
\alpha(q,\omega) = \frac{1}{12v_F^3 q} \left[(2\mu+\omega)^3 -3v_F^2 q^2 (2\mu+\omega) + 2 v_F^3 q^3 \right]
\label{Pi7}
\end{equation}
and
\begin{equation}
\beta(q,\omega) = \ln \left| \frac{2\mu + \omega - v_F q}{v_F q - \omega} \right| .
\label{Pi8}
\end{equation}
The full response function in the doped case is of course $\Pi_{\rho\sigma_z} = \Pi^\mathrm{in}_{\rho\sigma_z} +  \Pi^\mathrm{ex}_{\rho\sigma_z}$. This response function satisfies
\begin{equation}
\Pi_{\rho\sigma_z}(q \hat{\mathbf{z}},\omega)
  = \frac{\omega}{v_Fq}\, \Pi_{\rho\rho}(q \hat{\mathbf{z}},\omega) ,
\end{equation}
which relates it to the charge response.

\begin{figure}[tb]
\includegraphics[width=0.8\columnwidth]{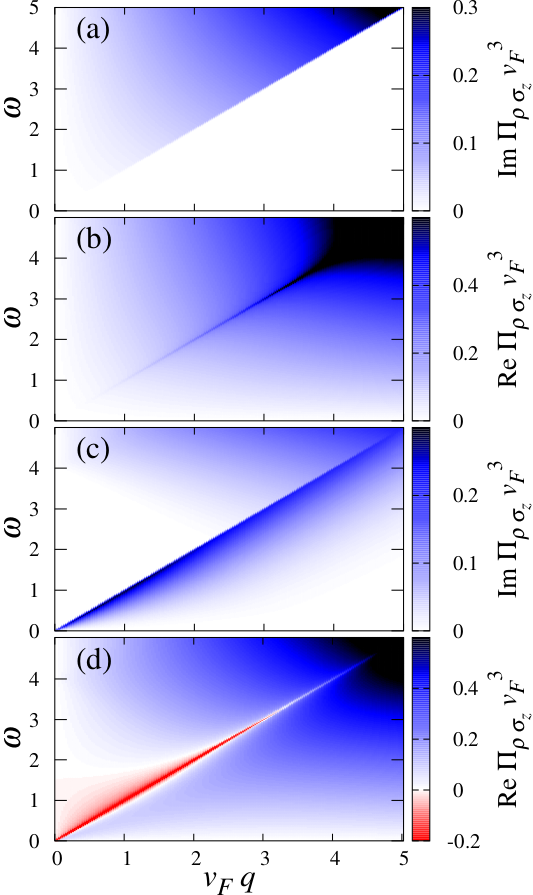}
\caption{Coupled charge-spin responses of a Weyl cone: wave-vector and frequency dependence of (a) $\mathrm{Im}\,\Pi_{\rho \sigma_z}$ and (b) $\mathrm{Re}\,\Pi_{\rho \sigma_z}$ for the undoped case ($\mu=0$). (c) $\mathrm{Im}\,\Pi_{\rho \sigma_z}$ and (d) $\mathrm{Re}\,\Pi_{\rho \sigma_z}$ for a doped ($\mu = 2.0$) Weyl cone.}
\label{fig_cs_den} 
\end{figure}

\begin{figure}[tb]
\includegraphics[width=0.7\columnwidth]{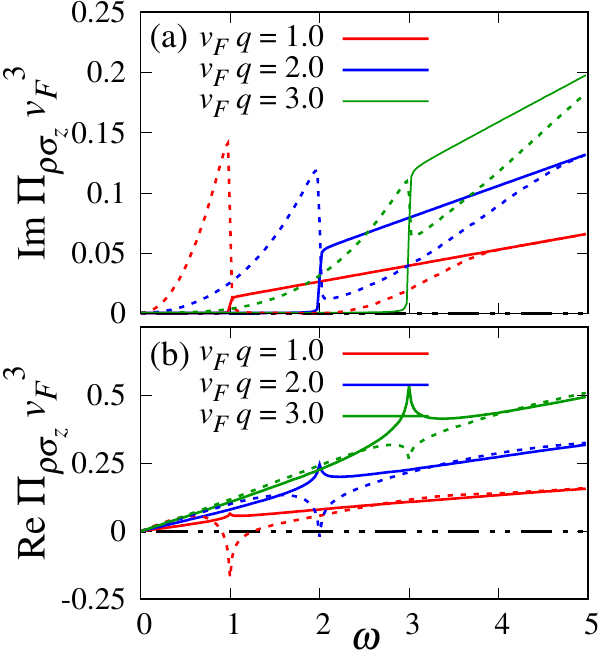}
\caption{Frequency dependence of (a) the imaginary and (b) the real part of the charge-spin response function $\Pi_{\rho\sigma_z}(q\hat{\mathbf{z}},\omega)$ for various values of the wave number $q$. Solid and dashed lines correspond to the undoped ($\mu = 0$) and the doped ($\mu = 1.5$) case, respectively.}
\label{fig_cs} 
\end{figure}

The wave-vector and frequency dependence of the longitudinal charge-spin response function $\Pi_{\rho\sigma_z}$ of a single Weyl cone with $\chi = +1$ is presented in Fig.\ \ref{fig_cs_den} for the undoped and doped cases. In Fig.\ \ref{fig_cs}, we plot the same data for cuts at fixed wave numbers. For the undoped case, when the valence band is completely occupied and the conduction band is empty, the imaginary part shows a step at $\omega = v_F q$, indicating the onset of interband particle-hole excitations (the finite slope of the step arises from the small imaginary part $\delta = $ 0.005 used for the calculations). The real part depends linearly on ${\bf q}$ and $\omega$ for small wave number or frequency. By dint of the Kramers-Kronig relations, the step in the imaginary part implies a logarithmic divergence (rounded by $\delta>0$) at $\omega = v_F q$ in the real part. Upon doping, a nonzero imaginary part appears for $v_F q - 2 \mu < \omega < v_F q$, resulting from intraband particle-hole excitations, while interband transitions are Pauli blocked for $v_F q < \omega < 2 \mu - v_F q$. Consequently, the step in the imaginary part and the peak in the real part become inverted and the real part can even change sign. Note that this does not imply a thermodynamic instability since the response of the system to external perturbations is governed by the full response tensor $\Pi_{\alpha\alpha'}$; an instability would be signaled by a negative real part of an eigenvalue.

It is worth noting that due to spin-momentum locking the charge-spin response is related to the density-current response calculated in Ref.\ \cite{Zhou2018}. The result for the density-current response for a single Weyl node is similar to the one obtained for the charge-spin response here. However, since the sign of spin-momentum locking is dictated by chirality, the density-current response is even in chirality, whereas the charge-spin response is odd, which has significant consequences as discussed in the next section.

The $4\times4$ response tensor $\Pi_{\alpha\alpha'}$ is related to the $6\times 6$ electromagnetic susceptibility by~\cite{Yarlagadda2000}
\begin{align}
\chi^{\rm ee}_{ij} &= \frac{\partial P_i}{\partial E_j}
  = - \frac{e^2}{q_i q_j}\, \Pi_{\rho \rho} , \\
\chi^{\rm mm}_{ij} &= \frac{\partial M_i}{\partial B_j}
  = \left( \frac{g \mu_B}{2} \right)^{\!2} \Pi_{\sigma_i \sigma_j} , \\
\chi^{\rm em}_{ij} &= \frac{\partial P_i}{\partial B_j}
  = i\, \frac{e g \mu_B}{2}\, \frac{1}{q_i}\, \Pi_{\rho \sigma_j} , \\
\chi^{\rm me}_{ij} &= \frac{\partial M_i}{\partial E_j}
  = i\, \frac{e g \mu_B}{2}\, \frac{1}{q_j}\, \Pi_{\sigma_i \rho} ,
\end{align}
where we have taken the electron charge to be $-e$ and its magnetic moment to be $-g \mu_B/2$. The charge-spin response tensor thus implies a magnetization response to an electric field and a polarization response to an electric field, as expected from the topological ${\bf E} \cdot {\bf B}$ term in the electromagnetic action \cite{Zyuzin2012}. Moreover, the fact that only the longitudinal charge-spin response is nonzero is consistent with the longitudinal nature of this term.

\section{Observable consequences}
\label{sec3}

In this section, we analyze how the charge-spin response is manifested in WSMs, which necessarily contain at least two Weyl nodes with vanishing total chirality. We will discuss the electromagnetic response and collective excitations.

\subsection{Electromagnetic response}
\label{sec3a}

The charge-spin response originates from the coupling of spin and momentum. Consequently, it changes sign for a Weyl cone with opposite chirality, $\chi = - 1$, in Eq.\ (\ref{Hamiltonian}). The response at low energies is obtained from our results in Sec.\ \ref{sec2} by adding the responses for each node. The measurable quantities are the electromagnetic susceptibilities $\chi_{ij}^\mathrm{em}(\mathbf{q},\omega)$ and $\chi_{ij}^\mathrm{me}(\mathbf{q},\omega)$. In WSMs with inversion symmetry (and broken TRS), any Weyl node at a momentum $\mathbf{k}$ is accompanied by another node at $-\mathbf{k}$ and the same energy with the opposite chirality. In this case, the total charge-spin response and also the off-diagonal spin-spin response must vanish in equilibrium.

On the other hand, in the presence of TRS (and broken inversion symmetry) any Weyl node at $\mathbf{k}$ is accompanied by another node at $-\mathbf{k}$ and the same energy with the \emph{same} chirality. The vanishing total chirality guarantees the existence of further nodes with opposite chirality but in general these are not required to have the same Fermi velocities or to lie at the same energy. Thus, the total charge-spin response need not cancel in equilibrium. Additional lattice symmetries can enforce the Weyl nodes to occur in pairs with the same Fermi velocities and energies but opposite chirality, in which case the response vanishes. The situation is similar if neither TRS nor inversion symmetry is present. In conclusion, in WSMs that break inversion symmetry, one has to analyze the remaining symmetries to ascertain whether the total charge-spin response vanishes in equilibrium. For example, we have found that for $(\mathrm{Gd},\mathrm{Nd})\mathrm{PtBi}$ in a magnetic field applied in a low-symmetry direction, no two Weyl nodes are at the same energy and a non-zero response is expected. We also note that for a range of field directions, there are exactly two Weyl points between the topmost two of the four low-energy bands.

Moreover, due to the chiral anomaly, the electron concentration in Weyl cones of opposite chirality is not conserved separately in parallel static $\mathbf{E}$ and $\mathbf{B}$ fields. The pumping of charge between Weyl nodes due to the anomaly is counterbalanced by inter-node scattering, leading to a non-equilibrium stationary state with different chemical potentials $\mu_\chi$ for Weyl nodes of different chirality $\chi$. (In WSMs with more than two Weyl nodes, the chemical potential will generally be distinct for all Weyl nodes and $\mu_n$ should be labeled by an index $n$ counting the nodes.) In such a case, the total charge-spin response will generally be nonzero. As an example, let us consider the ``hydrogen atom'' scenario for an inversion-symmetric WSM: two identical Weyl cones of opposite chirality at momenta $\pm\mathbf{k}_0$. The closest material example is the Dirac semimetal $\mathrm{Na}_3\mathrm{Bi}$, which becomes a WSM with two Weyl nodes in an applied static magnetic field \cite{Xiong2015}. In the presence of a parallel static electric field, the two nodes develop different effective chemical potentials due to the chiral anomaly. In this case, there will be a nonzero charge-spin response, which can be probed by measuring the electromagnetic susceptibilities for radio-frequency electric and magnetic fields.\footnote{The typical internode scattering time is $10^{-11}\,\mathrm{s}$} The dynamical charge-spin response  can thus serve as a probe of the Weyl character of quasiparticles.

\subsection{Collective excitations}
\label{sec3b}

Having investigated the linear charge-spin response of free Weyl fermions, we now turn to the collective excitations of a ``Weyl liquid'' in the presence of interactions. As noted in the previous section, the WSM must have low symmetry or be driven out of equilibrium to achieve nonzero charge-spin response. The interacting response functions are calculated within the random phase approximation (RPA). Since the transverse charge-spin response vanishes, the $4 \times 4$ RPA response tensor decomposes into two $2 \times 2$ blocks describing (i) the coupled charge and longitudinal spin response and (ii) the transverse spin response. Our interest is in the former part. We use a charge-spin basis, in which the coupled response takes the form
\begin{equation}
\hat\Pi = \left(\begin{array}{cc}
  \Pi_{\rho\rho} & \Pi_{\rho\sigma_z} \\
  \Pi_{\sigma_z\rho} & \Pi_{\sigma_z\sigma_z}
  \end{array} \right) .
\label{Matrix1}
\end{equation}
We first consider the consequences of an on-site interaction and then of long-range Coulomb repulsion. These two cases are understood as the extreme limits of the screened Coulomb interaction. We investigate the collective excitations for a single doped Weyl cone and then present results for a minimal model with two Weyl nodes of opposite chirality and different chemical potentials $\mu_\chi$. Details of the calculations are relegated to Appendix~\ref{app_collective}.

\subsubsection{On-site Hubbard interaction}
\label{sec3b1}

For an ordinary Fermi liquid with on-site Hubbard repulsion, the collective excitations in the low-temperature, collisionless regime consist of gapless zero-sound modes. We here consider a WSM with Hubbard repulsion of strength $U>0$ in the collisionless regime, taking $T=0$. Note that transport properties in the hydrodynamic regime have been studied by several authors~\cite{Gorbar2018a,Gorbar2018b,Lucas2016a}.

\begin{figure}[tb]
\includegraphics[width=0.8\columnwidth]{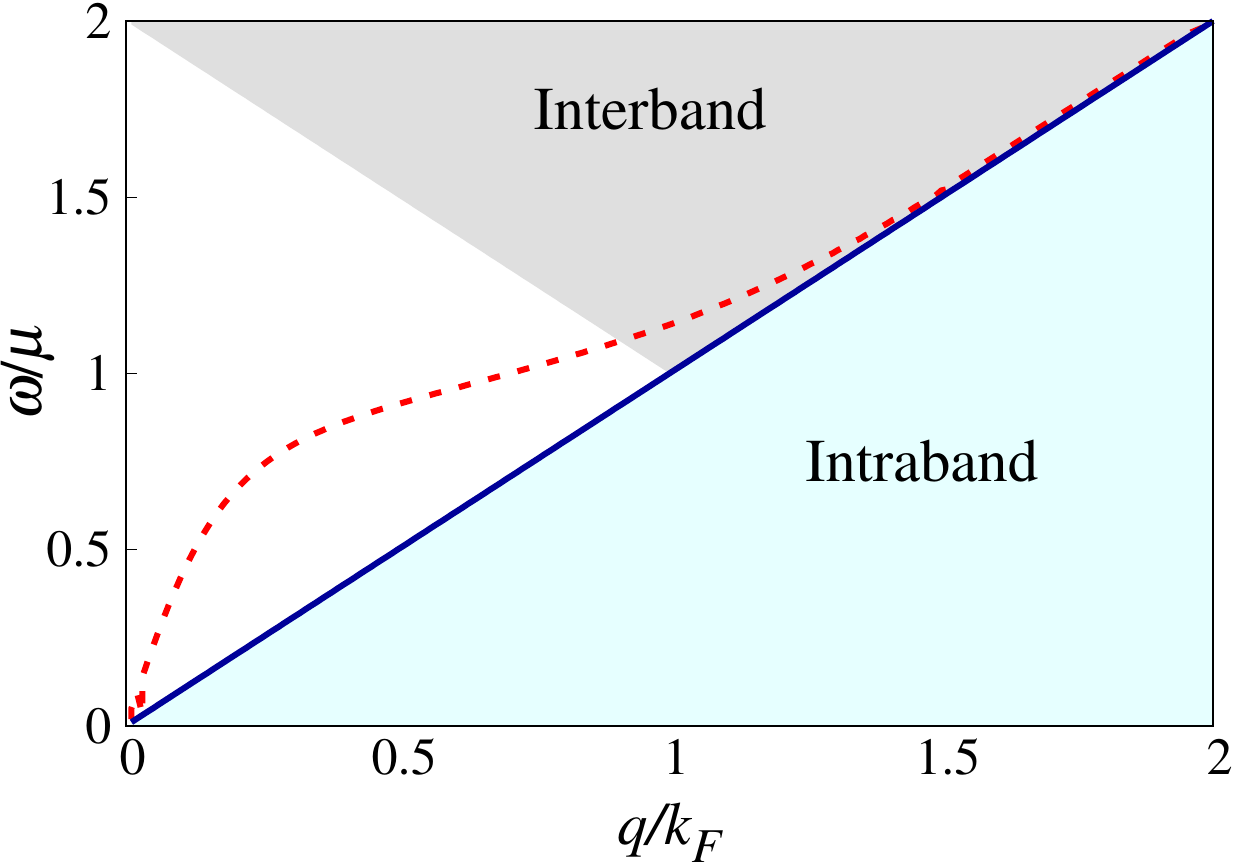}
\caption{Dispersion of collective modes for a single doped Weyl cone in the long-wavelength regime. The short-dashed red line shows the zero-sound mode without charge-spin coupling. In the presence of charge-spin coupling, the zero-sound mode linearly disperses with the Fermi velocity (solid blue line). Intraband particle-hole excitations are present in the light blue region for $\omega < v_F q$ and interband particle-hole excitations in the gray region with $\omega > v_F q$ and $\omega + v_F q > 2\mu$. Here, $U = v_F \Lambda = 10\mu$. For a typical bandwidth of $1\,\mathrm{eV}$, this corresponds to $U \approx 1\,\mathrm{eV}$ and $\mu \approx 0.1\,\mathrm{eV}$.} 
\label{fig_modes} 
\end{figure}

Using the matrix form of Eq.\ (\ref{Matrix1}), the RPA response is given by
\begin{equation}
\hat\Pi_\mathrm{RPA}({\bf q},\omega) = \hat\Pi({\bf q},\omega)\,
\left[\mathbbm{1} + U \tau_z \hat\Pi({\bf q},\omega)\right]^{-1} ,
\label{RPA_Hubbard}
\end{equation}
where $\mathbbm{1}$ is the unit matrix and $\tau_z$ is a Pauli matrix in the charge-spin basis. The latter implements the opposite sign of the Hubbard interaction in the charge and spin channels~\cite{Schrieffer1989}.

The poles of the eigenvalues of $\hat\Pi_\mathrm{RPA}({\bf q},\omega)$ determine the dispersion of collective modes. It is useful to contrast the results to the case without charge-spin coupling, $\Pi_{\rho \sigma_z} = 0$. In this case, the dispersion of the charge modes (zero sound) at small $q$ and $\omega$ is given by
\begin{equation}
\omega_\mathrm{zs}(q) = \sqrt{\cal A}\; \frac{2 \mu q\, [1 - \mathcal{O}(q^2)]}
  {\sqrt{1 + 2 {\cal A} q^2 \ln({v_F \Lambda}/{\mu})}} ,
\label{zero_sound}
\end{equation}
with ${\cal A} = {U}/{24\pi^2 v_F}$ (see Appendix \ref{app_collective}). The leading term for small $q$ is linear with velocity $2\mu \sqrt{\mathcal{A}}$. With increasing $q$, the velocity decreases and beyond a certain wave number $q \sim k_F$, the zero-sound mode enters the interband particle-hole continuum and becomes damped. Similarly to the ordinary Fermi liquid in the strong-interaction limit \cite{Pines1966}, the velocity of the zero sound is proportional to the square root of the interaction strength $U\propto \mathcal{A}$, and below a critical value $U_c = 6 \pi^2 v_F^3/\mu^2$, the zero sound gets damped by coupling to intraband particle-hole excitations (light blue region in Fig.~\ref{fig_modes}). On the other hand, the uncoupled longitudinal spin fluctuations would not support propagating collective modes since the Hubbard repulsion is effectively attractive in the spin channel.

\begin{figure}[tb]
\includegraphics[width=0.8\columnwidth]{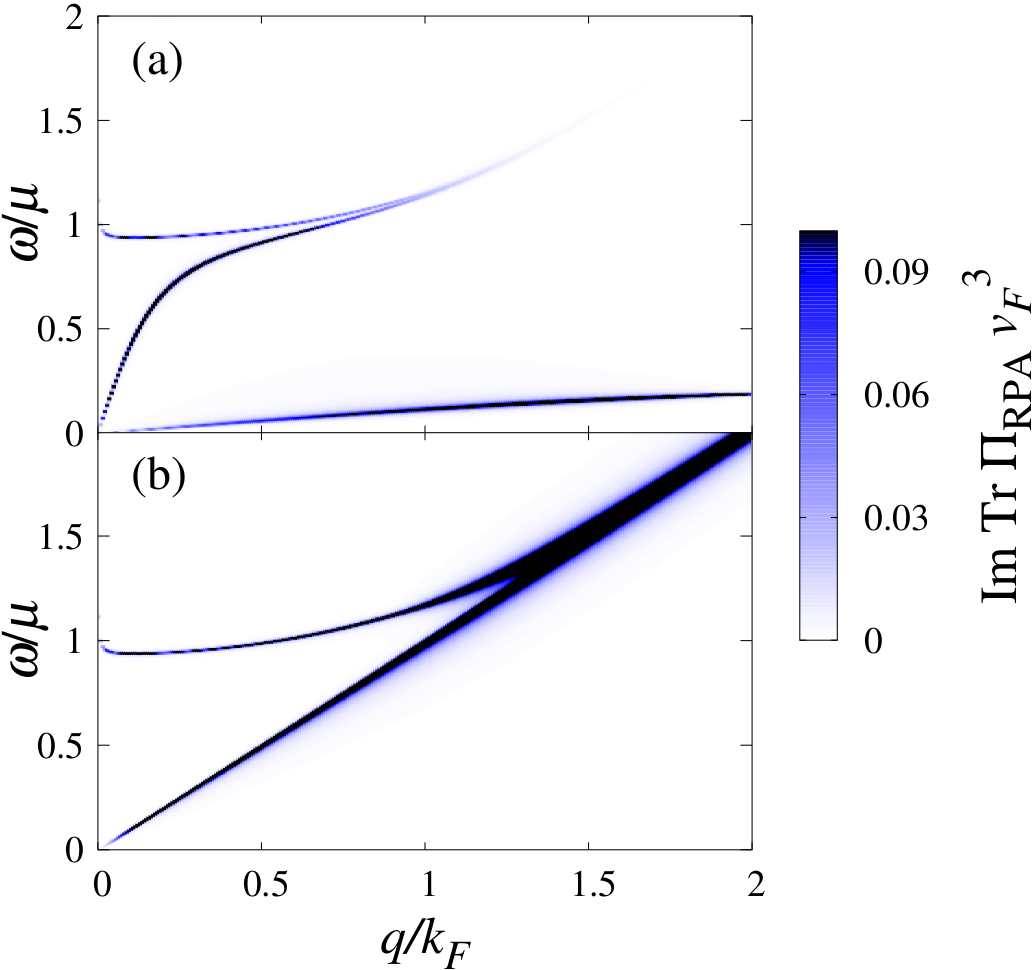}
\caption{Imaginary part of the trace of the RPA response function, $\mathrm{Im}\,\mathrm{Tr}\,\hat\Pi_\mathrm{RPA}(\mathbf{q},\omega)$ (a) without and (b) with charge-spin coupling. For parameter values see Fig.~\ref{fig_modes}.}
\label{fig_zs_spec} 
\end{figure}

In the presence of charge-spin coupling, the zero sound mixes with the longitudinal spin waves and the dispersion changes dramatically: it becomes lightlike, propagating with the Fermi velocity, as shown by the blue line in Fig.\ \ref{fig_modes}. The dispersion of this coupled zero-sound mode is \emph{critical} in the sense of having the minimum possible velocity for undamped excitations. Its velocity does not depend on the interaction strength or on the chemical potential, although the linewidth does. The collective modes can be clearly seen in the imaginary part of the RPA response function, $\mathrm{Im}\,\mathrm{Tr}\,\hat\Pi_\mathrm{RPA}(\mathbf{q},\omega)$, which is plotted in Fig.\ \ref{fig_zs_spec}. Aside from the zero-sound branch, which is strongly modified by charge-spin coupling, we observe a  gapped branch in Fig.\ \ref{fig_zs_spec}(a), which stems from pure spin fluctuations. The charge-spin coupling leads to its mixing with charge modes in Fig.\ \ref{fig_zs_spec}(b) but the dispersion remains qualitatively unchanged. The linearly dispersing branch at low frequencies in Fig.\ \ref{fig_zs_spec}(a) also consists of spin fluctuations. They are resonant with the intraband continuum and thus damped. In Fig.\ \ref{fig_zs_spec}(b), this branch is completely removed by the mixing with the charge modes.

What happens in a WSM with a pair of Weyl cones with opposite chirality that are otherwise identical? In equilibrium, the zero-sound mode is described only by the charge response and its dispersion will be qualitatively similar to the ordinary Fermi liquid as shown by the dotted line in Fig.\ \ref{fig_modes}. However, when charge is pumped from one Weyl node to the other by means of the chiral anomaly, the charge-spin response is turned on. As a result, the zero sound is described by the collective excitations of density and spin, as we will discuss now. Note that for materials of low symmetry, the charge-spin response will survive even in the absence of charge pumping.

Several proposals for the experimental identification of the chiral anomaly have been put forward, involving plasmons \cite{Lv2013,Zhou2015}, the optical conductivity \cite{Ashby2014,Goswami2015}, non-local transport measurements \cite{Parameswaran2014}, as well as others. Here, we show that zero sound can serve as a fingerprint of the chiral anomaly. In WSMs, the chiral anomaly leads to the transfer of charges from the $\chi = - 1$ nodes to $\chi = +1$, or vice versa, in parallel electric and magnetic fields. This process is balanced by internode scattering, characterized by a scattering time $\tau$, and a steady state with different charge densities in the two nodes is reached. The corresponding chirality-dependent chemical potential is given by \cite{Zhou2015,Thakur2018}
\begin{equation}
\mu_\chi = \mu\, \Big( 1 + \chi \nu^3 \Big)^{1/3} , 
\label{mu_chiral}
\end{equation}
with
\begin{equation}
\nu = \left( \frac{3e^2 v_F^3}{2 \mu^3}\, {\bf E} \cdot {\bf B}\, \tau \right)^{\!1/3}
\label{def_nu}
\end{equation}
being the ratio between pumped and equilibrium carrier densities.

\begin{figure}[tb]
\includegraphics[width=0.85\columnwidth]{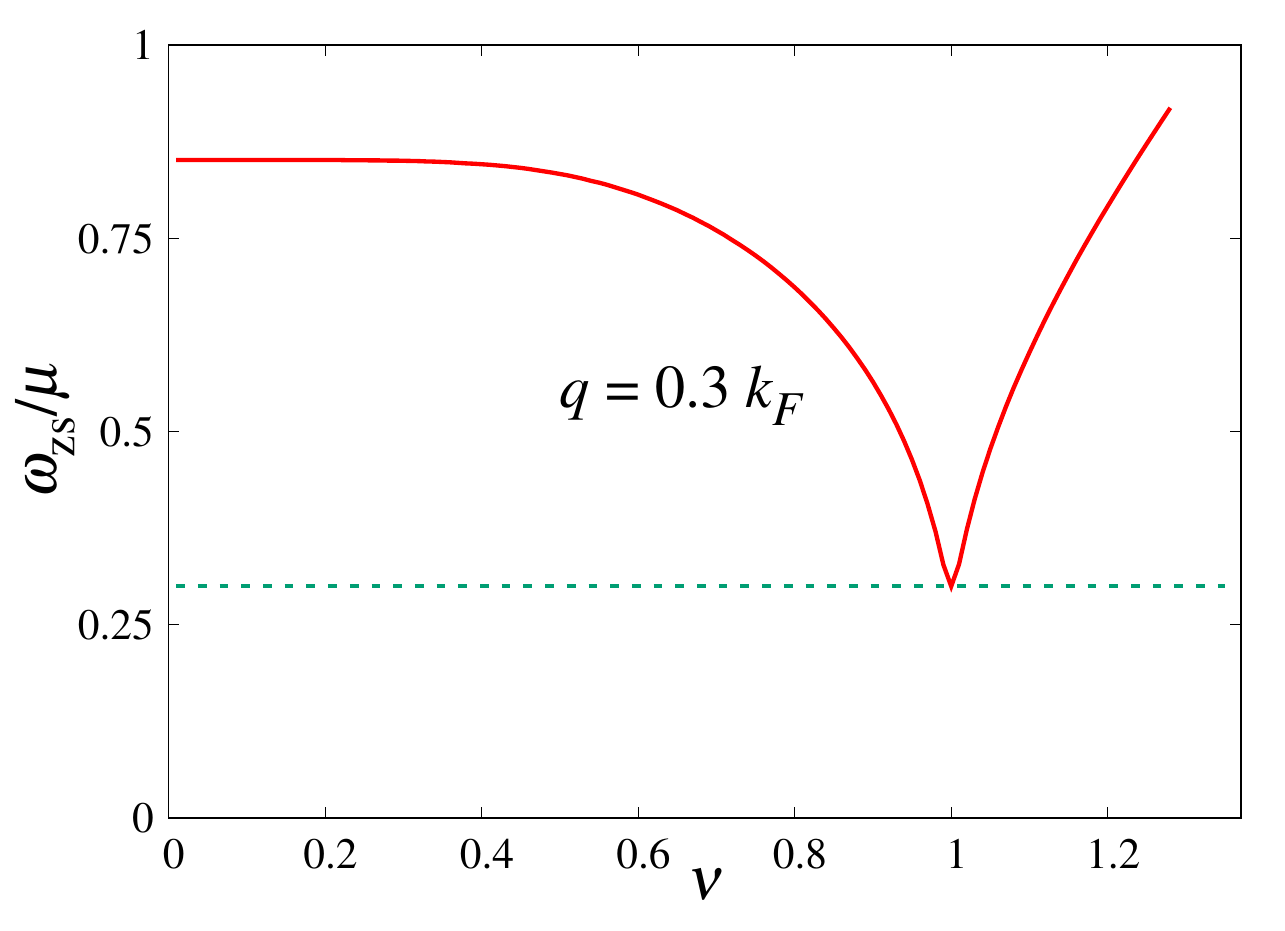}
\caption{Frequency of zero sound for a WSM with two Weyl nodes as a function of the ratio $\nu$ between pumped and equilibrium carrier densities (solid red line). The minimum occurs at the Lifshitz transition point $\nu = 1$, where the modes disperse with the Fermi velocity so that $\omega_\mathrm{zs} = v_Fq$ (dashed horizontal line). Parameters same as in Fig.~\ref{fig_modes}.}
\label{fig_chiral} 
\end{figure}

The dependence of the zero-sound frequency on chiral charge pumping is shown in Fig.\ \ref{fig_chiral}. In the absence of pumping, the positive and negative chiral nodes have the equilibrium chemical potential $\mu_+ = \mu_- = \mu$ (without loss of generality, we consider electron-doped Weyl cones, $\mu > 0$), and the zero sound is determined only by charge density fluctuations. As $\nu$ becomes nonzero by application of parallel static ${\bf E}$ and ${\bf B}$ fields, $\mu_+$ shifts upward while $\mu_-$ shifts downward. Consequently, the zero sound obtains contributions from spin fluctuations and is red-shifted (see Fig.\ \ref{fig_chiral}). At $\nu = 1$, $\mu_-$ crosses a Weyl node, which can be understood as a Lifshitz transition. At this transition, spin fluctuations have maximal contribution to the zero-sound wave and it becomes lightlike with velocity $v_F$. Beyond $\nu = 1$, when $\mu_-$ lies in the valence band, the frequency increases again. Therefore, the zero sound can be taken as a signature of the chiral anomaly, with a dip to lightlike dispersion indicating the Lifshitz transition point.

\subsubsection{Long-range Coulomb interaction}
\label{sec3b2}

\begin{figure}[tb]
\includegraphics[width=0.9\columnwidth]{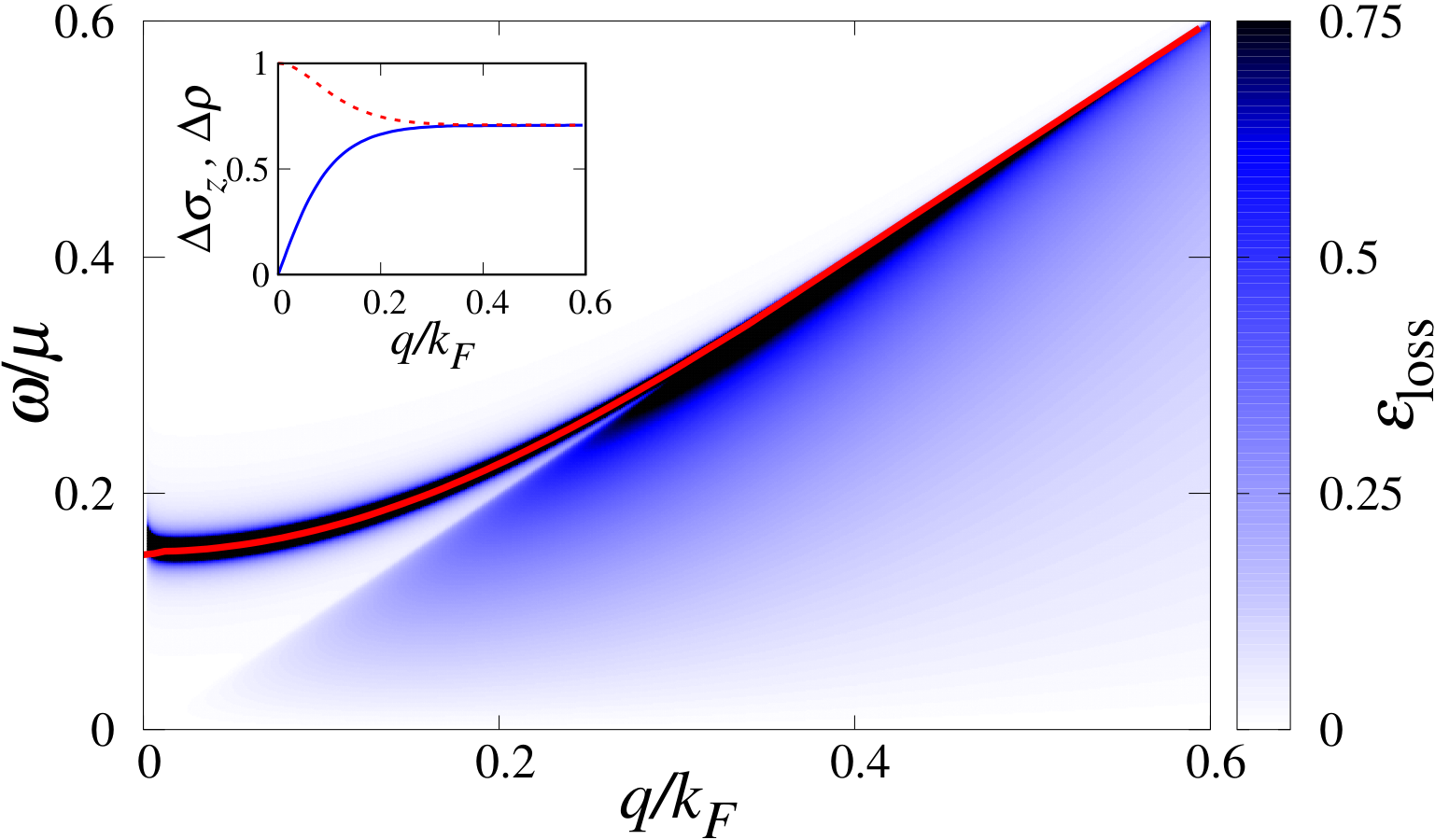}
\caption{Electron energy-loss function in the $(q,\omega)$ plane for a single Weyl cone. The red line denotes the spin-plasmon dispersion calculated from Eq.\ (\ref{plasmon}). The inset shows the amplitudes $\Delta\rho$ (solid blue line) and $\Delta\sigma_z$ (dashed red line) of the spin plasmon in the charge and spin channels, respectively. Here, we take $v_F \Lambda = 10 \mu$ and the background dielectric constant $\kappa = 20$.}
\label{fig_plasmon} 
\end{figure}

Now we consider Weyl fermions with a long-range Coulomb interaction. The RPA response is described by
\begin{equation}
\hat\Pi_\mathrm{RPA}({\bf q},\omega) = \hat\Pi({\bf q},\omega)\,
  \left[\mathbbm{1} + \frac{V(q)}{2}\, (\mathbbm{1} + \tau_z)\, \hat\Pi({\bf q},\omega)\right]^{-1}
\label{RPA_Coulomb}
\end{equation}
in the charge-spin basis. Here, $V(q) = 4 \pi e^2 / \kappa q^2$ is the Fourier-transformed Coulomb interaction, with $\kappa$ being the dielectric constant. The difference in the interaction vertex compared to Eq.\ (\ref{RPA_Hubbard}) arises from the fact that the Coulomb interaction only acts in the charge channel in the RPA, whereas the Hubbard interaction can be decomposed into charge and spin channels. Using that $\Pi_{\rho\rho}\Pi_{\sigma_z\sigma_z} - \Pi_{\rho\sigma_z}\Pi_{\sigma_z\rho} = 0$, the response function can be rewritten as
\begin{equation}
\hat\Pi_\mathrm{RPA}({\bf q},\omega) = \frac{1}{1 + V(q)\, \Pi_{\rho \rho}({\bf q},\omega)}\,
  \hat\Pi({\bf q},\omega).
\label{RPA_Coulomb2}
\end{equation}
Thus, similarly to the ordinary Fermi liquid \cite{Pines1966} and to the two-dimensional Dirac liquid \cite{Lucas2016b}, the zero sound morphs into plasmonic modes in the presence of long-range Coulomb interaction. Interestingly, unlike for a local interaction, all components of the charge-spin response tensor are uniformly enhanced at the RPA level and only the charge response governs this enhancement. The plasmon dispersion is thus given by the zeros of the RPA dielectric function \cite{Zhou2015}
\begin{equation}
\epsilon_{\rm RPA}({\bf q},\omega) = 1 + V(q)\, \Pi_{\rho \rho}({\bf q},\omega) ,
\label{plasmon}
\end{equation}
which also only depends on the charge response. The dispersion is thus the same as when charge-spin coupling is ignored, as in Ref.\ \cite{Zhou2015}. The dispersion can be obtained in the long-wavelength limit, keeping only the leading order in $q$ and is given by~\cite{Zhou2015}
\begin{equation}
\omega_{\rm pl}(q) = \omega_0  \left( 1 - \frac{v_F^2 q^2}{8 \mu^2}
  \left[ 1 +  \frac{\nu_0^2 - 3/5}{\nu_0^2\, (1-\nu_0^2)^2} \right] \right) ,
\end{equation}
where $\nu_0 = \omega_0/2\mu$ and
\begin{equation}
\omega_0 = \mu\, \sqrt{\frac{2 \alpha_{\kappa}}{3 \pi \kappa^*(\omega_0)}}
\end{equation}
is the plasmon frequency at $\mathbf{q}\rightarrow 0$. It is determined by $\alpha_{\kappa} = e^2/\kappa v_F$ and the frequency-dependent effective background dielectric function
\begin{equation}
\kappa^*(\omega) = 1 + \frac{\alpha_{\kappa}}{6\pi}\,
  \ln \left| \frac{4 v_F^2 \Lambda^2}{4 \mu^2 - \omega^2} \right| .
\end{equation}
The plasmons are manifested as sharp peaks in the electron energy-loss function
\begin{equation}
\mathcal{E}_{\rm loss}({\bf q},\omega) = - {\rm Im}\, \frac{1}{\epsilon_{\rm RPA}({\bf q},\omega)} ,
\end{equation}
which can be probed for example by electron energy-loss spectroscopy. Figure \ref{fig_plasmon} shows $\mathcal{E}_{\rm loss}$ in the $(q,\omega)$ plane for a single Weyl cone. The plasmon dispersion calculated from Eq.\ (\ref{plasmon}) is shown as the red line, which agrees well with the position of the peak in $\mathcal{E}_{\rm loss}$. The nonzero loss below the line $\omega = v_Fq$ stems from intraband particle-hole excitations. We see that the plasmon dispersion is gapped in the Weyl liquid, similarly to the ordinary Fermi liquid.

\begin{figure}[tb]
\includegraphics[width=0.8\columnwidth]{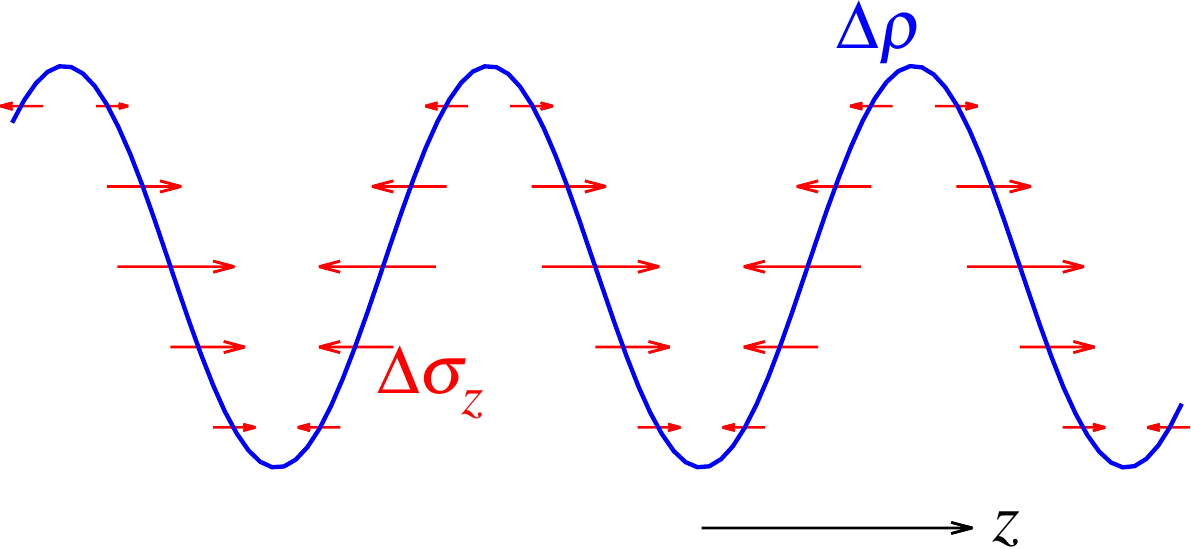} 
\caption{Schematic representation of a spin plasmon: a charge density wave (blue line) is mixed with a longitudinal spin polarization (red arrows). At the peaks and troughs of the density wave, the spin polarization vanishes, whereas at zeros of the density modulation, the spin polarization is maximum.}
\label{fig_spin_plasmon} 
\end{figure}

The fundamental difference between the collective modes in the Weyl liquid and in a normal Fermi liquid is that plasmons in the former carry spin, and hence can be called \textit{spin plasmons}. This is a signature of spin-orbital locking. A schematic representation of the spin plasmon is shown in Fig.\ \ref{fig_spin_plasmon}. A density fluctuation is accompanied by a $90^\circ$ out-of-phase longitudinal spin fluctuation. This phase shift and the amplitudes $\Delta\rho$ and $\Delta\sigma_z$ of the charge and spin fluctuations, respectively, can be calculated from the eigenvectors of the response tensor. They are plotted in the inset in Fig.\ \ref{fig_plasmon}. We find that the charge amplitude is linear in $q$ in the long-wavelength limit. Therefore, the spin fluctuation associated with spin plasmon becomes dominant for long wavelengths, and for $\mathbf{q}\rightarrow 0$ the spin plasmon becomes a pure spin excitation.

\begin{figure}[tb]
\includegraphics[width=0.9\columnwidth]{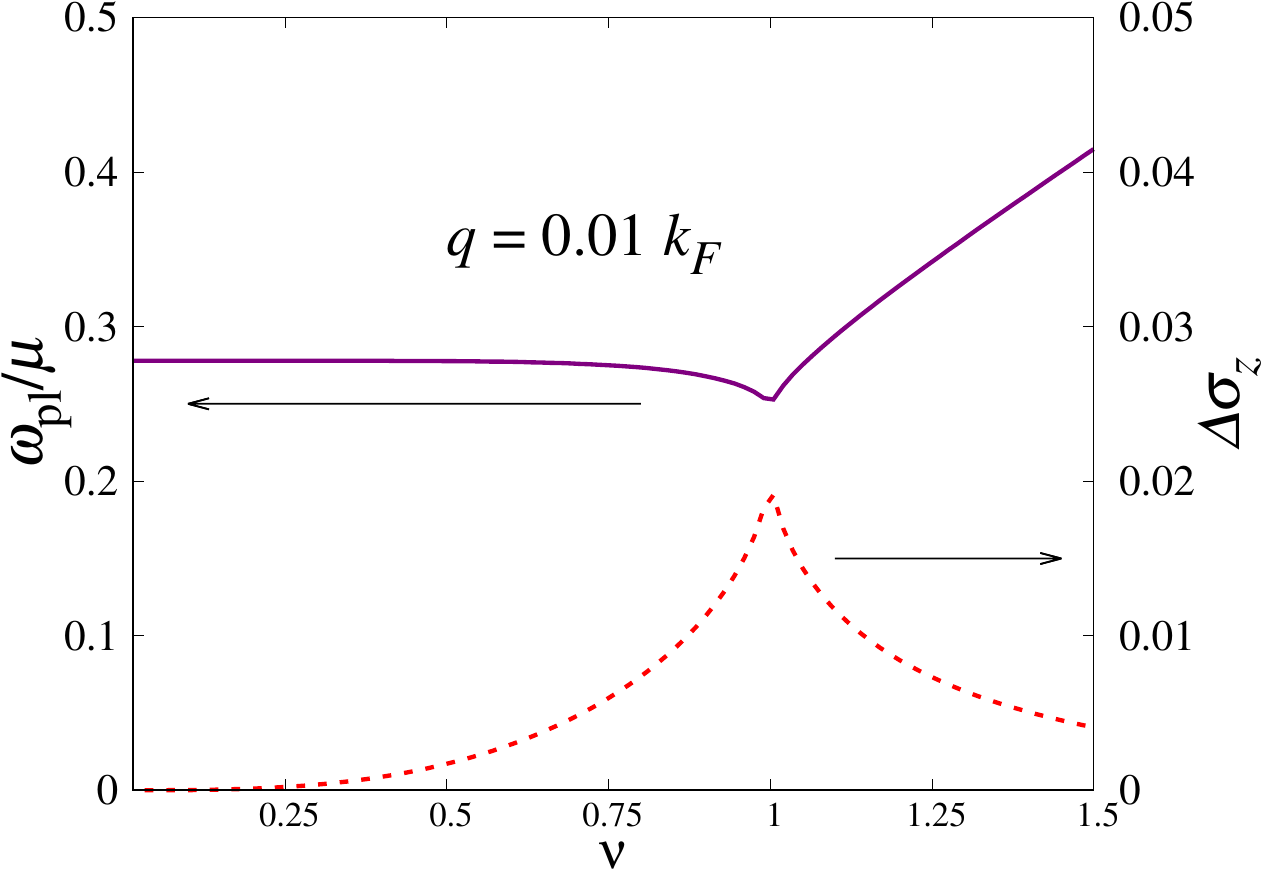}
\caption{Plasmon frequency (solid purple line) and the amplitude of the spin-wave part of the spin plasmon (dashed red line) for a pair of  Weyl nodes of opposite chirality as functions of the ratio $\nu$ between pumped and equilibrium carrier densities [see Eq.\ (\ref{def_nu})]. Here, $q = 0.01\, k_F$ and the other parameters are the same as in Fig.~\ref{fig_plasmon}.}
\label{fig_plasmon_freq}
\end{figure}

Spin plasmons have also been proposed for the two-dimensional helical liquid formed by the surface states of a topological insulator \cite{Raghu2010}. The main differences to our case are that (i) the spin plasmon of the topological insulator is a surface plasmon, whereas we are considering bulk excitations and (ii) the spin plasmon of the helical surface liquid carries transverse spin fluctuations, whereas the WSM spin plasmon is associated with longitudinal spin fluctuations.

We now turn to a WSM with a pair of Weyl cones with opposite chirality that are otherwise identical. In equilibrium (for ${\bf E} \cdot {\bf B} = 0$), the propagating plasmons do not carry any spin polarization. For ${\bf E} \cdot {\bf B}\neq 0$, the Weyl cones of opposite chirality have different chemical potentials, which results in the generation of a spin polarization by density fluctuations. The effect is shown in Fig.\ \ref{fig_plasmon_freq} as a function of charge pumping, denoted by $\nu$ [see Eq.\ (\ref{def_nu})]. The spin amplitude associated with the spin plasmon initially increases with $\nu$. At $\nu = 1$, the chemical potential $\mu_-$ crosses a Weyl node and the charge-spin coupling reaches its maximum, resulting in the maximum amplitude of the spin-wave component. Beyond $\nu = 1$, the spin-wave amplitude decreases again. Therefore, the spin wave associated with the collective mode can be used as a fingerprint of the position of the Weyl node. On the other hand, the spin-plasmon frequency decreases slightly as a function of $\nu$, reaches a minimum at the Lifsitz transition point $\nu = 1$, and then increases again~\cite{Zhou2015}.

It should be possible to detect the coupled charge-spin collective modes using optical pump-probe spectroscopy. In the following, we call the excitations ``spin plasmons'' but the discussion remains valid in the limit of short-range interactions as well. The basic idea is to generate a propagating longitudinal spin wave and to detect the charge density wave coupled to it. An intense circularly polarized light pulse propagating along the \textit{x} direction is incident on a \textit{yz} surface of the sample. The pulse generates a spin polarization, which is localized at the surface, with a typical length scale given by the optical penetration depth. This spin-polarization pattern can be decomposed into plane waves $e^{iqx}$. In the presence of charge-spin mixing, the spin waves dress with charge fluctuations to form spin plasmons, which then propagate through the bulk of the sample with group velocity $d\omega_\mathrm{pl}/dq$. The charge fluctuation associated with the spin plasmons could be detected at the opposite surface for example by means of the energy-gain (anti-Stokes) lines in inelastic light scattering. The dispersion $\omega_\mathrm{pl}(q)$ leads to a characteristic temporal distribution of the charge fluctuations at the probe surface, which could be accessed by varying the delay between pump and probe pulses. The amplitude of charge fluctuation depends on the value of ${\bf E \cdot B}$, and is maximum when the chemical potential for one chirality reaches a Weyl node.

Our scheme is quite different from the one proposed by Raghu \textit{et al.}\ \cite{Raghu2010} for the helical surface states of a topological insulator. There, a \emph{transverse} spin wave with given wave vector is coherently excited by the incident light. This approach does not work in our case since the spin wave needs to be longitudinal to couple to the plasmon.

\section{Summary and conclusions}
\label{sec4}

To summarize, we have investigated the electric and magnetic response of Weyl fermions by calculating all terms of the $4 \times 4$ charge-spin linear-response tensor for Weyl nodes with arbitrary doping. We have found that the charge and the longitudinal (i.e., parallel to the vector vector) spin responses are strongly coupled because of the spin-momentum locking in the Weyl Hamiltonian. The transverse spin response does not couple to the charge and longitudinal spin response to linear order. Our description of the Weyl nodes as separate ideal Weyl Hamiltonians limits the validity of our results to small momenta and low energies, where the appropriate momentum and energy scales are set by the deviation of the band structure from the ideal linear form.

With the full response tensor for a single Weyl node in hand, we have examined the electromagnetic response and the collective excitations of Weyl fermions in the presence of electron-electron interaction. The observation of charge-spin response requires either low crystal symmetry so as to avoid cancellation of contributions from Weyl nodes of opposite chirality or driving the system out of equilibrium. The latter can be achieved by applying parallel static electric and magnetic fields, by virtue of the chiral anomaly. For a local, Hubbard-type interaction, we obtain zero sound with a constant velocity given by the Fermi velocity $v_F$ of the Weyl fermions, independently of the interaction strength $U$ or the chemical potential $\mu$. This is a dramatic change compared to the result when charge-spin coupling is ignored; in that case, the dispersion is linear only in the long-wavelength limit and the velocity strongly depends on $U$ and $\mu$. The charge-spin coupling also causes the zero sound modes to have mixed charge and longitudinal spin character, and so gives rise to the possibility to ``hear'' the spin fluctuations, as proposed earlier for surface states of topological insulators~\cite{Raghu2010}. On the other hand, for a long-range Coulomb interaction, we have found a propagating spin plasmon, which also consists of charge and longitudinal spin fluctuations. While the character of the collective plasma modes is thus strongly affected by charge-spin coupling, their dispersion is still determined by the zeros of the dielectric function, which only depends on the charge response.

For real WSMs of high symmetry, the charge-spin coupling vanishes in equilibrium and the collective excitations are conventional. Therefore, the collective modes are sensitive probes of the chiral anomaly, which drives the system out of equilibrium in static parallel $\mathbf{E}$ and $\mathbf{B}$ fields. For a local interaction, the zero sound goes through a sharp dip to the critical frequency $v_F q$ when the chemical potential for one chirality is tuned to the Weyl point. On the other hand, for the Coulomb interaction, the spin content of the spin plasmons shows a sharp peak in this case. We have proposed a pump-probe experiment involving the spin plasmons that gives a null result in the absence of the chiral anomaly and yields the largest signal when the chemical potential for one chirality is tuned to the Weyl point. These signatures thus have the potential to lead to smoking-gun experiments for the chiral anomaly and the presence of Weyl nodes. As another possible direction for future research, the tunability of the spin plasmons by means of the chiral anomaly suggests the combination of plasmonics and spintronics in Weyl systems.

\vspace{2ex}
\acknowledgments

The authors thank T. Meng and A. Dey for useful discussions. Financial support by the Deutsche Forschungsgemeinschaft through project A04 of Collaborative Research Center SFB 1143 is gratefully acknowledged.

\onecolumngrid

\appendix

\section{Charge response}
\label{app_charge}

In this appendix, we reproduce the explicit form of the charge response tensor \cite{Lv2013,Zhou2015} for completeness. It is independent of chirality. The charge response can be written as the sum of a contribution from the undoped system (``intrinsic'') and a contribution due to doping (``extrinsic''),
\begin{align}
\Pi_{\rho\rho}(\mathbf{q},\omega)
  = \Pi_{\rho\rho}^\mathrm{in}(\mathbf{q},\omega)
  + \Pi_{\rho\rho}^\mathrm{ex}(\mathbf{q},\omega) .
\end{align}
The imaginary and real parts of the intrinsic contribution are given by
\begin{align}
{\rm Im}\,\Pi_{\rho\rho}^\mathrm{in}(\mathbf{q},\omega) &= \frac{q^2}{24 \pi v_F}\, \theta(\omega - v_F q) ,
\label{IPirhorhoin} \\
{\rm Re}\,\Pi_{\rho\rho}^\mathrm{in}(\mathbf{q},\omega) &= \frac{q^2}{24 \pi^2 v_F}\,
  \ln\left| \frac{4 v_F^2 \Lambda^2}{v_F^2 q^2 - \omega^2} \right| ,
\label{RPirhorhoin}
\end{align}
respectively, where $\theta(x)$ is the Heaviside step function. For electron doping, $\mu = v_F k_F > 0$, the imaginary and real parts of the extrinsic contribution read
\begin{align}
{\rm Im}\,\Pi^\mathrm{ex}_{\rho\rho}(\mathbf{q},\omega)
  &= \frac{1}{8 \pi v_F}\, \bigg[ \theta(v_F q - \omega) \big( [\alpha(q,\omega) - \alpha(q,-\omega)]
  \theta(2\mu - v_F q - \omega) + \alpha(q,\omega) \theta(2\mu - v_F q + \omega)
  \theta(v_F q + \omega - 2\mu) \big) \nonumber \\
&{}+ \theta(\omega - v_F q) \bigg( {-} \alpha(-q,-\omega) \theta(2\mu + v_F q - \omega)
  \theta(v_F q + \omega - 2\mu) - \frac{q^2}{3}\, \theta(2\mu - v_F q - \omega) \bigg) \bigg] , \\ 
{\rm Re}\,\Pi^\mathrm{ex}_{\rho\rho}(\mathbf{q},\omega)
  &= \frac{1}{8\pi^2 v_F} \bigg[ \frac{8\mu^2}{3v_F^2} - \alpha(q,\omega)\beta(q,\omega)
  - \alpha(-q,\omega)\beta(-q,\omega) - \alpha(q,-\omega)\beta(q,-\omega)
  - \alpha(-q,-\omega)\beta(-q,-\omega) \bigg] ,
\label{RPirhorhoex}
\end{align}
where we have defined
\begin{equation}
\alpha(q,\omega) = \frac{1}{12v_F^3 q} \left[(2\mu+\omega)^3 -3v_F^2 q^2 (2\mu+\omega) + 2 v_F^3 q^3 \right]
\label{def_alpha}
\end{equation}
and
\begin{equation}
\beta(q,\omega) = \ln \left| \frac{2\mu + \omega - v_F q}{v_F q - \omega} \right| .
\label{def_beta}
\end{equation}

\section{Spin response}
\label{app_spin}

Here, we discuss the calculation of the $3 \times 3$ spin response tensor. The components can be written as
\begin{equation}
\Pi_{\sigma_l \sigma_m} ({\bf q},i\omega_n)
  = - \frac{1}{N} \sum_{\bf k} \sum_{\lambda,\lambda'} \langle \phi_{\lambda'}({\bf k} + {\bf q})|
    \sigma_l |\phi_{\lambda}({\bf k}) \rangle\, \langle \phi_{\lambda}({\bf k})|
    \sigma_m |\phi_{\lambda'}({\bf k} + {\bf q})\rangle\,
    \frac{n^F_{\lambda}({\bf k}) - n^F_{\lambda'}({\bf k} + {\bf q})}{i\omega_n + \epsilon_{\lambda}({\bf k}) - \epsilon_{\lambda'}({\bf k} + {\bf q})} ,
\label{Pi_ss1}
\end{equation}
where $\lambda, \lambda'=\pm$ refer to the two bands with energies $\epsilon_\pm = \pm v_F |{\bf k}|$ and eigenvectors (periodic parts of Bloch states) given by
\begin{align}
|\phi_{+}({\bf k})\rangle &= \left( \begin{array}{c}
  \cos\frac{\theta_\mathbf{k}}{2}\, e^{-i \varphi_\mathbf{k}/2} \\[0.7ex]
  \sin\frac{\theta_\mathbf{k}}{2}\, e^{i \varphi_\mathbf{k}/2}
  \end{array} \right) , \\
|\phi_{-}({\bf k})\rangle &= \left( \begin{array}{c}
  \sin\frac{\theta_\mathbf{k}}{2}\, e^{-i \varphi_\mathbf{k}/2} \\[0.7ex]
  - \cos\frac{\theta_\mathbf{k}}{2}\, e^{i \varphi_\mathbf{k}/2}
  \end{array}\right) ,
\end{align}
where $\theta_\mathbf{k}$ and $\varphi_\mathbf{k}$ are the polar and azimuthal angle of ${\bf k}$, respectively. $n_\lambda^F$ is the Fermi-Dirac distribution function for the band $\lambda$, which becomes a step function for the zero-temperature limit considered here.

The response tensor has diagonal ($i = j$) and off-diagonal ($i \ne j$) terms. The interband and intraband contributions to diagonal terms can be simplified to \cite{Thakur2018}
\begin{equation}
\Pi_{\sigma_l \sigma_l}^\mp({\bf q},\omega) = \frac{1}{2N} \sum_{\bf k}
  \bigg(1 \pm \frac{k'_m k_m + k'_n k_n - k'_l k_l}{k' k} \bigg)
  \bigg[\frac{1}{v_F (k \pm k') - \omega - i \delta}
    + \frac{1}{v_F ( k \pm k') + \omega + i \delta} \bigg] ,
\label{Pi_ss2}
\end{equation}
where again ${\bf k'} = {\bf k} + {\bf q}$ and $l$, $m$, $n$ refer to three orthogonal coordinate axes with $\varepsilon_{lmn} = + 1$.  

Taking $\mathbf{q} = q\hat{\mathbf{z}}$, the diagonal components can be categorized into longitudinal and transverse responses. After rewriting the momentum sum as an integral and performing the angular part, the longitudinal part becomes
\begin{equation}
\Pi_{\sigma_z \sigma_z}^\mp(q \hat{\mathbf{z}},\omega) = \pm \frac{1}{16 \pi^2 q^3} \int dk \int dk'\,
  (k \pm k')^2\, [q^2 - (k\mp k')^2]
  \left[ \frac{1}{v_F (k \pm k') - \omega - i \delta} + (\omega \rightarrow - \omega) \right] ,
\label{Pi_ss3}
\end{equation}
while the transverse response is described by
\begin{align}
\Pi_{\sigma_{x} \sigma_{x}}^\mp(q \hat{\mathbf{z}},\omega)
  &= \Pi_{\sigma_{y} \sigma_{y}}^\mp(q \hat{\mathbf{z}},\omega) \nonumber \\
&= \frac{1}{32 \pi^2 q^3} \int dk \int dk'\, [(k + k')^2 \mp q^2]\, [q^2 \pm (k - k')^2]\,
  \left[ \frac{1}{v_F (k \pm k') - \omega - i \delta} + (\omega \rightarrow - \omega) \right] .
\label{Pi_ss4}
\end{align}
Here, the limits of the integrals over $k$ and $k'$ depend on the chemical potential in such a way that only transitions from occupied to empty states are included. The integrals also depend on an ultraviolet cutoff, which is necessary since the Hamiltonian in Eq.\ (\ref{Hamiltonian}) describes an infinite sea of negative-energy states~\cite{Principi2009,Sabio2008}.

The diagonal spin responses for a single Weyl cone were calculated by Thakur \textit{et al.}\ \cite{Thakur2018} and Zhou and Chang \cite{Zhou2018}. It is even in chirality $\chi$. The real parts contain a term ${\Lambda^2}/{6 \pi^2 v_F}$, which depends on the ultraviolet cutoff $\Lambda$ \cite{Polini2009,Kargarian2015,Thakur2018,Zhou2018,Ominato2018}. The spin response function is regularized by taking the $\mu = 0$ ground state as the reference \cite{Sabio2008,Polini2009,Thakur2018}, which amounts to subtracting the $\Lambda^2$ term.

The result for the intrinsic (undoped) contributions can be expressed in terms of the charge response as
\begin{align}
\Pi_{\sigma_z \sigma_z}^\mathrm{in}(q\hat{\mathbf{z}},\omega) &= \frac{\omega^2}{v_F^2 q^2}\,
  \Pi_{\rho \rho}^\mathrm{in}(q\hat{\mathbf{z}},\omega) ,
\label{Pizzin} \\
\Pi_{\sigma_x \sigma_x}^\mathrm{in}(q\hat{\mathbf{z}},\omega) &= \frac{\omega^2 - v_F^2 q^2}{v_F^2 q^2}\,
  \Pi_{\rho \rho}^\mathrm{in}(q\hat{\mathbf{z}},\omega) .
\label{Pixxin}
\end{align} 
However, for the extrinsic (doping) part, the relationship between the spin and charge response is only valid for the longitudinal response function,
\begin{equation}
\Pi_{\sigma_z \sigma_z}^\mathrm{ex}(q\hat{\mathbf{z}},\omega)
  = \frac{\omega^2}{v_F^2 q^2}\, \Pi_{\rho \rho}^\mathrm{ex}(q\hat{\mathbf{z}},\omega) ,
\end{equation}
whereas the transverse spin response does not satisfy a simple relation to the charge response \cite{Thakur2018}. The imaginary part of the extrinsic transverse response function reads as~\cite{Thakur2018}
\begin{align}
\mathrm{Im}\,\Pi_{\sigma_x \sigma_x}^\mathrm{ex}(q\hat{\mathbf{z}},\omega)
&= \frac{\omega^2 - v_F^2 q^2}{32 \pi v_F^3 q^3}\, \bigg[ \theta(v_F q - \omega)
  \big( [\gamma(q,\omega) - \gamma(q,-\omega)] \theta(2\mu - v_F q - \omega) \nonumber \\
&{}+ \gamma(q,\omega) \theta(2\mu - v_F q + \omega) \theta(v_F q + \omega - 2\mu) \big) \nonumber \\
&{}+ \theta(\omega - v_F q) \bigg(  \gamma(-q,-\omega) \theta(2\mu + v_F q - \omega)
  \theta(v_F q + \omega - 2\mu) + \frac{4q^2}{3}\, \theta(2\mu - v_F q - \omega) \bigg) \bigg] ,
\end{align}
where the function $\gamma(q,\omega)$ is defined as
\begin{equation}
\gamma(q,\omega) = 2q\, \alpha(q,\omega) + q^2\, (2 \mu - v_F q + \omega) ,
\end{equation}
with $\alpha(q,\omega)$ is defined in Eq.~(\ref{def_alpha}). The real part is given by~\cite{Thakur2018}
\begin{align}
\mathrm{Re}\,\Pi_{\sigma_x \sigma_x}^\mathrm{ex}(q\hat{\mathbf{z}},\omega)
&= - \frac{\omega^2 - v_F^2 q^2}{2v_F^2 q^2}\, \Pi_{\rho \rho}^\mathrm{ex}(q\hat{\mathbf{z}},\omega)
  - \frac{\omega^2 - v_F^2 q^2}{16\pi^2 v_F^3 q} \sum_{\eta = \pm 1} \mathcal{P}\!
  \int_{0}^{k_F} dk\, \mathcal{P}\! \int_{|k-q|}^{k+q} dk'\, \bigg[
   \frac{v_F}{v_F k' + \eta v_F k + \omega} + (\omega \rightarrow - \omega) \bigg] \nonumber \\
&{}- \frac{1}{16 \pi^2 v_F q^3} \sum_{\eta = \pm 1} \int_{0}^{k_F} dk
  \int_{|k-q|}^{k+q} dk'\, (k' - \eta k) \big[ (k' + \eta k)^2 + q^2 \big] .
\label{Pixxex1}
\end{align}
We have re-evaluated the integrals since Eq.\ (D4) in Ref.\ \cite{Thakur2018} contains an ambiguous factor $0/0$ for $c=0$ and is incorrect if one naively cancels $c$ before setting $c=0$. Our result reads as
\begin{align}
\mathrm{Re}\,&\Pi_{\sigma_x \sigma_x}^\mathrm{ex}(q\hat{\mathbf{z}},\omega)
= - \frac{\omega^2 - v_F^2 q^2}{2v_F^2 q^2}\,
  \Pi_{\rho \rho}^\mathrm{ex}(q\hat{\mathbf{z}},\omega)
  - \frac{\mu^2}{4 \pi^2 v_F^3} \nonumber \\
&{} - \frac{\omega^2 - v_F^2 q^2}{32\pi^2 v_F^4 q}\,
  \Bigg( \theta(v_Fq-\mu)
    \big[ \xi(q,\omega) \beta(-q,\omega) + \xi(q,-\omega) \beta(-q,-\omega)
    - \xi(-q,\omega) \beta(q,\omega) - \xi(-q,-\omega) \beta(q,-\omega) \big] \nonumber \\
&{} + \theta(\mu-v_Fq) \bigg[ (2\mu+\omega) \ln \left| \frac{\xi(q,\omega)}{\xi(-q,\omega)} \right|
    + (2\mu-\omega) \ln \left| \frac{\xi(q,-\omega)}{\xi(-q,-\omega)} \right|
    - 2 \omega \ln \left| \frac{v_Fq + \omega}{v_Fq - \omega} \right|
     \nonumber \\
&\quad {}+ v_Fq\, \big[ \zeta(q,\omega) + \zeta(-q,\omega)
      + \zeta(q,-\omega) + \zeta(-q,-\omega) \big] \bigg] \Bigg) ,
\end{align}
where $\xi(q,\omega) = 2\mu + v_F q + \omega$ and
\begin{equation}
\zeta(q,\omega) = \ln\left| \frac{2\mu + v_Fq + \omega}{v_Fq + \omega} \right| .
\end{equation}
This also simplifies the expressions since the last term in Eq.\ (\ref{Pixxex1}) is found to reduce to the momentum- and frequency-independent term $-\mu^2/4\pi^2v_F^3$.

Note that the full longitudinal response function, $\Pi_{\sigma_z\sigma_z} = (\omega^2/v_F^2 q^2)\, \Pi_{\rho \rho}$, has the same sign as the charge response function \cite{Thakur2018,Zhou2018}. This is different from the linear response of the helical surface states of a topological insulator, where the \emph{transverse} spin response is related to the charge response and the relative factor is $-\omega^2/v_F^2 q^2$~\cite{Raghu2010}.

The interband and intraband contributions to the off-diagonal spin responses can be obtained from
\begin{align}
\Pi_{\sigma_l \sigma_m}^\mp({\bf q},\omega)
  &= \frac{1}{2N} \sum_{\bf k} \bigg( \left[ {\mp} \frac{k_l k'_m + k'_l k_m}{k k'}
  + i \left(\frac{k'_n}{k'} \pm \frac{k_n}{k}\right) \right]
  \frac{1}{v_F ( k \pm k') - \omega - i \delta} \nonumber \\
&{}+ \left[ {\mp} \frac{k_l k'_m + k'_l k_m}{k k'}
  - i \left(\frac{k'_n}{k'} \pm \frac{k_n}{k}\right) \right]
  \frac{1}{v_F (k \pm k') + \omega + i \delta} \bigg] .
\label{Pi_ss5}
\end{align}
Still assuming $\mathbf{q}=q\hat{\mathbf{z}}$, we have $\Pi_{\sigma_x \sigma_z} = \Pi_{\sigma_y \sigma_z} = 0$ and after performing the angular integrals, Eq.\ (\ref{Pi_ss5}) simplifies to
\begin{equation}
\Pi^\mp_{\sigma_x \sigma_y}(q \hat{\mathbf{z}},\omega)
  = \frac{i}{16 \pi^2 q^2} \int dk \int dk'\, (k \mp k')\,
  [q^2 - (k \pm k')^2] \left[ \frac{1}{v_F (k \pm k') - \omega - i \delta}
  - (\omega \rightarrow - \omega) \right] .
\label{Pi_ss6}
\end{equation}
The integration limits again depend on the chemical potential in such a way that only transitions from occupied to empty states are included. The response can be decomposed into intrinsic and extrinsic parts. Due to spin-momentum locking, it is proportional to the off-diagonal current-current correlation function, which in turn is related to the Hall conductivity~\cite{Raghu2010}, giving
\begin{equation}
\sigma_{xy}(\mathbf{q},\omega)
  = \frac{i e^2 v_F^2}{\omega}\, \Pi_{\sigma_x \sigma_y}(\mathbf{q},\omega) .
\label{HallB1}
\end{equation}
As pointed out by Burkov and Balents \cite{Burkov2011b}, an overall constant in the Hall conductivity cannot be determined within a continuum model. This problem carries over to the off-diagonal transverse spin response. In the evaluation of Eq.\ (\ref{Pi_ss5}) or Eq.\ (\ref{Pi_ss6}), it appears as an ambiguity of how to regularize the cutoff-dependent term. Specifically, for a lattice model, the sum over $\mathbf{k}$ in Eq.\ (\ref{Pi_ss5}) depends on the location of the Weyl node in momentum space. In the static and uniform limit, Burkov and Balents \cite{Burkov2011b} determine this offset from the known limit of vanishing anomalous Hall effect in a trivial insulator. This leads to the Hall conductance being proportional to the separation of nodes \cite{Yang2011,Burkov2011b,Zyuzin2012,Kargarian2015}. The generalization of this approach to the frequency- and wave-vector-dependent response is difficult since it would involve the regularization of a cutoff-dependent \emph{function} of $\mathbf{q}$ and $\omega$. Since the transverse spin (or Hall) response is not central for this paper, we do not attempt this here. Accordingly disregarding the cutoff-dependent term, the intrinsic part is found to be purely imaginary,
\begin{equation}
\Pi_{\sigma_x \sigma_y}^{\rm in}(q \hat{\mathbf{z}},\omega) = i\, \frac{q \omega}{24 \pi^2 v_F^2} .
\end{equation}
For the extrinsic contribution, a lengthy derivation yields
\begin{align}
{\rm Im}\,\Pi_{\sigma_x \sigma_y}^{\rm ex}(q \hat{\mathbf{z}},\omega)
  &= \frac{1}{16\pi^2 v_F^5 q^2}\, \bigg[ \frac{\omega^2 - v_F^2 q^2}{4}\,
  \bigg( \xi(-q,\omega) \xi(q,\omega)  \ln \left|\frac{\xi(-q,\omega)}{\xi(q,\omega)}\right|
  + \xi(q,-\omega) \xi(-q,-\omega)  \ln \left|\frac{\xi(q,-\omega)}{\xi(-q,-\omega)}\right|
  \nonumber \\
&{}+ \big[\xi(q,\omega) \xi(-q,\omega) + \xi(q,-\omega) \xi(-q,-\omega)\big]
  \ln \left| \frac{v_F q + \omega}{v_F q - \omega} \right| \bigg)
  + 4 v_F q \omega\, (\mu^2 - v_F^2 q^2) \bigg] , \\
{\rm Re}\,\Pi_{\sigma_x \sigma_y}^{\rm ex}(q \hat{\mathbf{z}},\omega)
  &= \frac{v_F^2 q^2 - \omega^2}{64\pi^2 v_F^5 q^2}\, \big[ \xi(q,\omega) \xi(-q,\omega)
  \theta(v_F q - \omega) \theta(2\mu - v_F q + \omega) \nonumber \\
&{}- \xi(q,-\omega) \xi(-q,-\omega) \theta(\omega - v_F q)
  \theta(2\mu + v_F q - \omega) \theta(v_F q + \omega - 2\mu) \big] ,
\end{align}
where $\xi(q,\omega) = 2\mu + v_F q + \omega$. The transverse off-diagonal spin response is odd in chirality~\cite{Zhou2018}. We note that the results of Zhou and Chang \cite{Zhou2018} are equivalent to the ones of Thakur \textit{et al.}\ \cite{Thakur2018} and those obtained here, except for a slightly different form of the cutoff-dependent term.

\section{Coupled charge-spin response}
\label{app_coupled}

In this appendix, we discuss the main steps of the calculation of the charge-spin response for a single Weyl cone with $\chi = +1$. By virtue of spin-momentum locking, it is proportional to the density-current response, which was calculated in Ref.\ \cite{Zhou2018} using the Passarino-Veltman reduction scheme \cite{PaV79}. The charge-spin response function can be calculated from the density-spin correlations as
\begin{equation}
\Pi_{\rho \sigma_l} ({\bf q},i\omega_n)
  = - \frac{1}{N} \sum_{\bf k} \sum_{\lambda,\lambda'} \langle\phi_{\lambda'}({\bf k} + {\bf q})|
  \phi_{\lambda}({\bf k})\rangle\, \langle\phi_{\lambda}({\bf k})|\sigma_{l}
  |\phi_{\lambda'}({\bf k} + {\bf q})\rangle\,
  \frac{n^F_{\lambda}({\bf k}) - n^F_{\lambda'}({\bf k} + {\bf q})}
  {i\omega_n + \epsilon_{\lambda}({\bf k}) - \epsilon_{\lambda'}({\bf k} + {\bf q})}
\label{Pi_cs1}
\end{equation}
(see Appendix \ref{app_spin} for definitions of symbols). At zero temperature, the interband (superscript $-$) and intraband (superscript $+$) contributions simplify to
\begin{align}
\Pi_{\rho \sigma_l}^\mp({\bf q},\omega)
  &= \frac{1}{2N} \sum_{\bf k} \bigg[ \bigg(\frac{k'_l}{k'} \mp \frac{k_l}{k}
  \pm i \frac{k_m k'_n - k'_m k_n}{k k'}\bigg)\,
  \frac{1}{v_F ( k \pm k') - \omega - i \delta} \nonumber \\
&{}+ \bigg(\frac{k_l}{k} \mp \frac{k'_l}{k'} \pm i\, \frac{k_m k'_n - k'_m k_n}{k k'}\bigg)\,
  \frac{1}{v_F ( k \pm k') + \omega + i \delta} \bigg] .
\label{Pi_cs2}
\end{align}
Further calculation shows that the transverse components of the charge-spin response, $\Pi_{\rho \sigma_l}^\mp(\mathbf{q},\omega)$ for $\hat{\bf l} \perp {\bf q}$, vanish. We can thus write the vector of the response functions $\Pi_{\rho\sigma_l}^\pm$ as
\begin{equation}
\Pi_{\rho {\boldsymbol \sigma}}^\mp({\bf q},\omega)
  = \Pi_{\rho \sigma_q}^\mp({\bf q},\omega)\, \hat{\mathbf{q}}
  = \Pi_{\rho \sigma_z}^\mp(q\hat{\mathbf{z}},\omega)\, \hat{\mathbf{q}} ,
\end{equation}
where $\hat{\mathbf{q}} \equiv \mathbf{q}/q$ is the unit vector in the direction of $\mathbf{q}$ and $\sigma_q \equiv {\boldsymbol \sigma}\cdot\hat{\mathbf{q}}$.

Taking $\mathbf{q} = q\hat{\mathbf{z}}$, without loss of generality, the non-vanishing response functions can be written in terms of the polar angles of wave vectors as
\begin{equation}
\Pi_{\rho \sigma_z}^\mp(q\hat{\mathbf{z}},\omega)
  = \frac{1}{2N} \sum_{\bf k} (\cos \theta_{\bf k'}  \mp \cos \theta_{\bf k})
    \left[\frac{1}{v_F ( k \pm k') - \omega - i \delta} - (\omega \rightarrow - \omega) \right] .
\label{Pi_cs3}
\end{equation}
Due to particle-hole symmetry, the charge-spin response is identical for $\mu \rightarrow - \mu$, and we now consider an electron-doped Weyl node, i.e., $\mu > 0$. The response can be decomposed into a sum of intrinsic (undoped) and extrinsic (doping) parts, $\Pi_{\rho \sigma_z} = \Pi^{\rm in}_{\rho \sigma_z} + \Pi^{\rm ex}_{\rho \sigma_z}$. The intrinsic part only contains interband contributions, whereas the extrinsic part can be divided into interband and intraband contributions. We first evaluate the intrinsic part, which is given by
\begin{equation}
\Pi_{\rho \sigma_z}^{\rm in}(q\hat{\mathbf{z}},\omega)
  = \frac{1}{16 \pi^2 q^2} \int_{0}^{\Lambda} dk \int_{|k-q|}^{k+q} dk'\, (k+k')\, [q^2 - (k-k')^2]
  \left[ \frac{1}{v_F(k+ k') - \omega  - i \delta} - (\omega \rightarrow - \omega) \right] ,
\label{Pi_cs4}
\end{equation}
where we have made the limits of integration explicit. $\Lambda$ is the ultraviolet cutoff for the momentum integral. Using the Sokhotski-Plemelj formula
\begin{equation}
\frac{1}{x\pm i\epsilon} = \mathcal{P}\, \frac{1}{x} \mp i \pi \delta(x)
\end{equation}
and taking $\omega > 0$, we write
\begin{align}
{\rm Im}\,\Pi_{\rho \sigma_z}^{\rm in}(q\hat{\mathbf{z}},\omega)
  &= \frac{1}{16 \pi q^2 v_F} \int_{0}^{\Lambda} dk \int_{|k-q|}^{k+q} dk'\, (k+k')\, [q^2 - (k-k')^2]\,
  \delta(\tilde{\omega} - k - k') , \\
{\rm Re}\,\Pi_{\rho \sigma_z}^{\rm in}(q\hat{\mathbf{z}},\omega)
  &= \frac{1}{16 \pi^2 q^2 v_F}\, \mathcal{P}\! \int_{0}^{\Lambda} dk \int_{|k-q|}^{k+q} dk'\,
  (k+k')\, [q^2 - (k-k')^2]
  \left( \frac{1}{k + k' - \tilde{\omega}} - \frac{1}{k + k' + \tilde{\omega}} \right) ,
\label{Pi_cs5}
\end{align}
with $\tilde{\omega} = \omega/v_F$. The imaginary part is easy to calculate,
\begin{equation}
{\rm Im}\,\Pi_{\rho \sigma_z}^{\rm in}(q\hat{\mathbf{z}},\omega)
  = \frac{q \omega}{24 \pi v_F^2}\, \theta(\omega - v_F q).
\label{Pi_cs6}
\end{equation}
The real and imaginary parts are related by causality. However, since the imaginary part diverges for $\omega  \rightarrow \infty$, the standard Kramers-Kronig relation is not applicable. A generalized Kramers-Kronig relation can be used, though. The \textit{n}th-order generalized Kramers-Kronig relation for a response function $\chi(\omega)$ that diverges as $\omega^{n-1}$ for $\omega \rightarrow \infty$ is given by \cite{Bjorken1965,Thakur2018}
\begin{equation}
\frac{{\rm Re}\,\chi(\omega)}{\omega^n}
  = \frac{1}{\omega}\, \lim_{\zeta \to 0} \frac{{\rm Re}\,\chi(\zeta)}{\zeta^{n-1}}
  + \frac{1}{\pi}\, \mathcal{P}\! \int_{-\infty}^{\infty} d\zeta\,
  \frac{{\rm Im} \,\chi(\zeta)}{\zeta^n\,(\zeta - \omega)} .
\label{Pi_cs7}
\end{equation}
Here, the imaginary part of the charge-spin response diverges linearly with $\omega$, thus we employ the second-order Kramers-Kronig relation, which yields
\begin{equation}
{\rm Re}\,\Pi_{\rho \sigma_z}^{\rm in}(q\hat{\mathbf{z}},\omega)
  = \omega \lim_{\zeta \to 0} \frac{{\rm Re}\,\Pi_{\rho \sigma_z}^{\rm in}(q\hat{\mathbf{z}},\zeta)}{\zeta}
  + \frac{\omega^2}{\pi}\, \mathcal{P}\! \int_{-\infty}^{\infty} d\zeta\,
  \frac{{\rm Im}\,\Pi_{\rho \sigma_z}^{\rm in}(q\hat{\mathbf{z}},\zeta)}{\zeta^2\,(\zeta - \omega)} .
\label{Pi_cs8}
\end{equation}
The first term on the right-hand side evaluates to
\begin{equation}
\frac{q \omega}{24 \pi^2 v_F^2}\, \left( \ln\frac{4\Lambda^2 - q^2}{q^2}
  - \frac{1}{2} \left[ \frac{(2\Lambda)^3}{q}\, \ln \frac{2\Lambda + q}{2\Lambda - q}
  - 2(2\Lambda)^2 - 6q\Lambda \ln \frac{2\Lambda + q}{2\Lambda - q} + \frac{16}{3}q^2 \right] \right) .
\label{Pi_cs9}
\end{equation}
Using the identities
\begin{align}
\lim_{x \to \infty} \bigg( \frac{x^3}{y}\, \ln \frac{x+y}{x-y} - 2x^2 \bigg) &= \frac{2y^2}{3} , \\
\lim_{x \to \infty} x \ln \frac{x+y}{x-y} &= 2y
\label{Pi_cs10}
\end{align}
in the limit of large cutoff $\Lambda$, the terms in the angular bracket in Eq.\ (\ref{Pi_cs9}) vanish, and the expression simplifies to 
\begin{equation}
\frac{q\omega}{24 \pi^2 v_F^2}\, \ln \frac{4\Lambda^2}{q^2}.
\label{Pi_cs11}
\end{equation}
The second term on the right-hand side of Eq.\ (\ref{Pi_cs8}) can be rewritten as
\begin{equation}
\frac{2\omega^2}{\pi} \int_0^{\infty} d\zeta\, \frac{\omega}{\zeta^2\,(\zeta^2 - \omega^2)}\,
  {\rm Im}\,\Pi_{\rho \sigma_z}(q\hat{\mathbf{z}},\zeta) = \frac{q\omega}{24 \pi^2 v_F^2}\,
  \ln \left| \frac{v_F^2 q^2}{v_F^2 q^2 - \omega^2} \right| .
\label{Pi_cs12}
\end{equation}
Therefore, the real part of the intrinsic charge-spin response is given by
\begin{equation}
{\rm Re}\,\Pi_{\rho \sigma_z}^{\rm in} (q\hat{\mathbf{z}},\omega) = \frac{q\omega}{24 \pi^2 v_F^2}\,
  \ln \left| \frac{4 v_F^2 \Lambda^2}{v_F^2 q^2 - \omega^2} \right| .
\label{Pi_cs13}
\end{equation}
Direct evaluation of the integral in Eq.\ (\ref{Pi_cs5}) gives the same result.

The extrinsic contribution for electron doping can be calculated in a similar fashion. The imaginary and real part are given by
\begin{align}
{\rm Im}\,\Pi^{\rm ex}_{\rho \sigma_z}(q\hat{\mathbf{z}},\omega)
  &= \frac{\omega}{8 \pi q v_F^2}\, \bigg( \theta(v_F q - \omega)\theta(2\mu - v_F q - \omega)\,
  [\alpha(q,\omega) - \alpha(q,-\omega)] \nonumber \\
&{}+ \theta(v_F q - \omega)\theta(2\mu - v_F q + \omega)\theta(v_F q + \omega - 2\mu)
  \alpha(q,\omega) \nonumber \\
&{}+ \theta(\omega - v_F q)\theta(2\mu + v_F q - \omega)\theta(v_F q + \omega - 2\mu)
  [{-} \alpha(-q,-\omega)]
+ \theta(\omega - v_F q)\theta(2\mu - v_F q - \omega) \left(- \frac{q^2}{3}\right) \! \bigg) ,
\label{Pi_cs14} \\
{\rm Re}\,\Pi^{\rm ex}_{\rho \sigma_z}(q\hat{\mathbf{z}},\omega)
  &= \frac{\omega}{8\pi^2 v_F^2 q}\, \left[ \frac{8\mu^2}{3v_F^2}
  - \alpha(q,\omega)\beta(q,\omega) - \alpha(-q,\omega)\beta(-q,\omega)
  - \alpha(q,-\omega)\beta(q,-\omega) - \alpha(-q,-\omega)\beta(-q,-\omega) \right] ,
\end{align}
where the functions $\alpha$ and $\beta$ are defined in Eqs.\ (\ref{def_alpha}) and (\ref{def_beta}), respectively. The step functions represent boundaries in the $(q,\omega)$ plane for interband and intraband particle-hole excitations. The charge-spin response functions are odd in chirality.

\section{Dispersion of collective modes}
\label{app_collective}

Here, we calculate the dispersion relation of the coupled collective modes. The real part of the charge response function $\Pi_{\rho\rho}$ \cite{Lv2013,Zhou2015} has intrinsic and extrinsic parts given in Eqs.\ (\ref{RPirhorhoin}) and (\ref{RPirhorhoex}), respectively. In the long-wavelength limit with $v_F q \ll \omega \ll 2 \mu$, the real part of $\Pi_{\rho\rho}$ to order $q^4$ is
\begin{equation}
\Pi_{\rho \rho}(\mathbf{q}, \omega) \cong \frac{q^2}{24\pi^2 v_F}
  \left[ \ln \frac{4v_F^2 \Lambda^2}{4\mu^2 - \omega^2}
  - \frac{4\mu^2}{\omega^2}
    \left( 1 - \frac{v_F^2 q^2}{4\mu^2}\, [1 + \mathcal{K}(\omega/2\mu)] \right) \right] ,
\label{Pi_cc1}
\end{equation}
with
\begin{equation}
\mathcal{K}(u) = \frac{u^2 - 3/5}{u^2(1- u^2)^2} .
\end{equation}
For a local Hubbard interaction $U$, the dispersion of zero sound is determined by $1 + U \, {\rm Re}\, \Pi_{\rho \rho}(\mathbf{q}, \omega_\mathrm{zs}(q)) = 0$ so that
\begin{equation}
\omega_\mathrm{zs}(q) \cong \sqrt{\frac{U}{24\pi^2 v_F}}\:
  \frac{2 \mu q}
  {\sqrt{ 1 + \frac{U}{24\pi^2 v_F}\, q^2 \ln\frac{4 v_F^2 \Lambda^2}{4 \mu^2 - \omega^2} }}
  \left(1 - \frac{v_F^2 q^2}{8\mu^2}\, [1 + \mathcal{K}(\omega/2\mu)] \right) .
\label{Pi_cc2}
\end{equation}
In the long-wavelength limit, the dispersion becomes
\begin{equation}
\omega_\mathrm{zs}(q) \cong \sqrt{\frac{U}{24\pi^2 v_F}}\: 2\mu q .
\end{equation}
Strictly speaking, Eq.\ (\ref{Pi_cc2}) should be solved self-consistently to obtain the dispersion. However, since the frequency satisfies $\omega \ll 2\mu$, we can ignore $\omega^2$ in the denominator of the logarithmic term and also take the long-wavelength form $\omega_0$ for calculating $\mathcal{K}$.

For the Coulomb interaction $V(q) = 4 \pi e^2 / \kappa q^2$, the dispersion of the spin plasmon is determined by $1 + V(q)\, {\rm Re}\, \Pi_{\rho \rho}(q, \omega) = 0$, which in the long-wavelength limit gives \cite{Zhou2015}
\begin{equation}
\kappa^*(\omega) - \frac{4 \mu^2 \alpha_{\kappa}}{6 \pi \omega^2}
  \left( 1 - \frac{v_F^2 q^2}{4\mu^2}\, \left[1 + \mathcal{K}(\omega/2\mu)\right] \right) \cong 0 ,
\end{equation}
with the effective background dielectric function
\begin{equation}
\kappa^*(\omega) = 1 + \frac{\alpha_{\kappa}}{6\pi}\,
  \ln \left| \frac{4 v_F^2 \Lambda^2}{4 \mu^2 - \omega^2} \right| .
\end{equation}
Hence, the dispersion reads as
\begin{equation}
\omega_\mathrm{pl}(q) \cong \mu\, \sqrt{\frac{2 \alpha_{\kappa}}{3 \pi \kappa^*(\omega)}}\,
  \left( 1 - \frac{v_F^2 q^2}{8\mu^2}\, [1 + \mathcal{K}(\omega/2\mu)] \right) .
\label{Pi_cc3}
\end{equation}
Again, Eq.\ (\ref{Pi_cc3}) should in principle be solved self-consistently but we can use the $q=0$ plasma frequency 
$\omega_0 = \mu\, \sqrt{{2 \alpha_{\kappa}}/{3 \pi \kappa^*(0)}}$ for calculating $\mathcal{K}$ and $\kappa^*(\omega)$ with reasonable accuracy. 

\twocolumngrid

\bibliography{Ghosh}

\end{document}